\documentclass[12pt,preprint]{aastex}

\newcommand{\et}{et al.}
\newcommand{\kms}{km s$^{-1}$}
\newcommand{\ha}{H$\alpha$}
\newcommand{\solar}{\ifmmode_{\sun}\;\else$_{\sun}\;$\fi}

\newcommand{\HII}{H$\,${\sc ii}}
\newcommand{\HI}{H$\,${\sc i}}

\newcommand{\x}{\enspace}
\newcommand{\xx}{\enspace\enspace}

\newcommand{\rtf}{R$_{25}$}

\newcommand{\rh}{R$_H$}

\newcommand{\sigcrit}{$\Sigma_{c}$}
\newcommand{\siggas}{$\Sigma_g$}
\newcommand{\sfrunit}{M\solar\ yr$^{-1}$ kpc$^{-2}$}
\newcommand{\sigha}{$\Sigma_{H\alpha}$}

\slugcomment{To appear in ApJ.}
\shorttitle{Broad-Band Imaging of Irregular Galaxies}
\shortauthors{Hunter and Elmegreen}

\begin{document}

\title{Broad-Band Imaging of a Large Sample of Irregular Galaxies}

\author{Deidre A. Hunter}
\affil{Lowell Observatory, 1400 West Mars Hill Road, Flagstaff, Arizona
86001 USA}
\email{dah@lowell.edu}

\and

\author{Bruce G. Elmegreen}
\affil{IBM T. J. Watson Research Center, PO Box 218, Yorktown Heights,
New York 10598 USA}
\email{bge@watson.ibm.com}

\begin{abstract}
We present the results of UBV imaging of a large sample of
irregular galaxies: 94 Im systems, 24 Blue Compact Dwarfs (BCDs),
and 18 Sm galaxies. We also include JHK imaging of 41 of these 
galaxies. The sample spans a large range in galactic parameters. 
Ellipse fit axial ratios, inclinations, and position angles 
are derived, integrated photometry and azimuthally-averaged 
surface photometry profiles are determined,
and exponential fits give the central surface brightnesses, 
scale lengths, and isophotal and half-power radii. These 
data are used to address the shapes of Im galaxies, look for 
clues to past interactions in large-scale peculiarities, examine the 
nature and consequences of bars, study color gradients and 
large-scale color variations, and compare the exponential 
disk profiles of the young and old stellar components. 
For example, color
gradients exhibit a great variety and not all passbands are correlated.
Bars are associated with higher star formation rates. Many irregulars
show a double exponential radial light profile that is steeper in the
outer parts, and these are reproduced by a new model of star formation
that is discussed in a companion paper.
Some galaxies, primarily BCDs, have double exponentials that
are steeper (and bluer) in the inner parts, presumably from centralized
star formation.  Im-types have thicker, less-prominent dust layers than
spiral galaxies because of their lower average surface densities and
midplane extinctions.
\end{abstract}


\section{Introduction}

Dwarf irregular (dIm) galaxies serve as laboratories of star
formation without the influence of spiral density waves or shear.
They have evolved relatively slowly over time and chemically
resemble the outer parts of present-day spirals. Ultra-low surface
brightness dIms may represent the slowest evolving galaxies in the
Universe. If dIms evolve via periodic starbursts, then they could
dominate intermediate redshift surveys. They are so numerous and
contain such a large fraction of their mass in gas, that the most
distant dwarfs may dominate the faint QSO absorption lines. They
are like the proposed building blocks of spiral galaxies in the
cold dark matter theory. For all of these reasons, dIm galaxies
are an important component of the Universe.

Normal dIm galaxies span a range of a factor of $10^4$ in star
formation rate per unit area, yet detailed studies of their
stellar populations suggest that most evolve with a star formation
rate that varies by only a factor of a few over time (Ferraro \et\
1989; Tosi \et\ 1991; Greggio \et\ 1993; Marconi \et\ 1995;
Gallart \et\ 1996a,b; Aparicio \et\ 1997a,b; Dohm-Palmer \et\
1998, Gallagher \et\ 1998, Gallart \et\ 1999). Some show evidence
of higher amplitude variations, however (Israel 1988, Tolstoy
1996, Dohm-Palmer \et\ 1997, Greggio \et\ 1998, Tolstoy \et\
1998).  What distinguishes dwarf galaxies with high star formation
rates from those with low rates? We believe this variation, which
is also present to a lesser extent in spirals (Kennicutt 1989;
Rocha-Pinto et al.\ 2000), indicates there is a fundamental aspect
of star formation that is missing from the current theory.

What governs star formation locally within a
galaxy is also not clear.  Dense, star-forming clouds and \HII\ regions form
near the central regions, yet the detectable stars extend far
beyond this, with the gas going out further still. Does star
formation also occur in the outer disk but without showing prominent
\HII\ regions?

In a differentially-rotating disk there is a critical column
density \sigcrit\ above which the disk is unstable to ring-like
perturbations in the radial direction (Toomre 1964). Kennicutt
(1989) and Martin \& Kennicutt (2001) determined the analogous
column density for star formation using a sample of Sc spiral
galaxies. They found that the ratio of observed gas density
\siggas\ to critical gas density \sigcrit\ has a characteristic
value at the radius where significant star formation ends.
Presumably gas is too stable to form stars further out.

Motivated by the success of the Toomre gravitational instability
model for spiral galaxies, we and others applied it to dIm
galaxies (Hunter \& Plummer 1996; Meurer \et\ 1996; van Zee \et\
1997a; Hunter, Elmegreen, \& Baker 1998).  The results suggested
that dIms should not be forming stars at all: most have
\siggas$<$\sigcrit\ throughout. The model also fails to predict
where star formation ends or which galaxies have higher rates. A
related result was found for spiral galaxies: stars continue to
form beyond the radius where the gas surface density drops below
the threshold (Ferguson et al. 1998). These results implied that
processes other than large-scale spontaneous instabilities are
important in tiny galaxies and the outer parts of spiral galaxies,
leading us to examine more local effects. These include triggering
from supernovae and other stellar pressures and gravitational
collapse following turbulence compression (see review by Mac Low
\& Klessen 2004).  Also likely to be important is the thermal
state of the gas, that is, whether cool clouds can form at the local
pressure and radiation field (e.g., Elmegreen \& Parravano 1994;
Wolfire et al. 2003; Schaye 2004).

In a study of the dIm NGC 2366 (Hunter, Elmegreen, \& van Woerden
2001a), we found that star formation occurs primarily where the gas
column density exceeds 6 M\solar\ pc$^{-2}$ (also see van der Hulst
\et\ 1993; van Zee \et\ 1997a; Meurer \et\ 1998). This threshold
corresponds approximately to the presence of a cool phase of \HI,
which may be a
second requirement for star formation. We also found that the peak
densities in regions of star formation are equal to the local
tidal densities for gravitational self-binding of a co-rotating
cloud against galactic tidal forces. This binding condition may be
more fundamental than the Toomre condition because it is local,
3-dimensional, and does not involve ring or spiral arm generation
as an intermediate step toward star formation.

To address the question of what regulates star formation in tiny
galaxies, we have conducted a multi-wavelength survey of a large
sample of reasonably normal, relatively nearby
galaxies without spiral arms. The data consist of UBV
and \ha\ images for the entire sample, and JHK images, \HI\
maps, CO observations, and \HII\ region spectrophotometry for a sub-sample.
The \ha, UBV, and JHK image sets act as probes of star formation on three
different times scales: \ha\ images
trace the most recent star formation ($\leq$10 Myrs) through
the ionization of natal clouds by the short-lived massive stars;
UBV, while a more complicated clue, is dominated by the stars formed
over the past Gyr for on-going star formation (Gallagher \et\ 1984);
and JHK integrates over the lifetime of the galaxy
(Hunter \& Gallagher 1985a).

So far, we have used these data to conduct
several case studies
(NGC 2366: Hunter, Elmegreen, \& van Woerden 2001a;
DDO 88: Simpson, Hunter, \& Knezek 2005a;
DDO 43: Simpson, Hunter, \& Nordgren 2005b)
as well as studies of the \HII\ region luminosity function and
distributions
(Youngblood \& Hunter 1999, Roye \& Hunter 2000),
gas abundances
(Hunter \& Hoffman 1999),
pressures of \HII\ regions relative to the background galactic disk
(Elmegreen \& Hunter 2000),
and V-band and \ha\ power spectra (Willett, Elmegreen, \& Hunter 2005).
The \ha\ data of the entire sample were presented
previously (Hunter \& Elmegreen 2004).
Here we present the UBVJHK imaging data of the full sample.

\section{The Sample} \label{sec-sample}

The 136 sample galaxies are listed in Table \ref{tab-sample} where
we have grouped the galaxies into three categories: Im---94 systems,
Blue Compact Dwarf (BCD)---24 systems, and Sm---18 systems.
The morphological classifications were taken from de Vaucouleurs
\et\ (1991$=$RC3).
Irregular galaxies were first described by Hubble (1926) as
``lacking both dominating nuclei and rotational symmetry'' with
the ``Magellanic Clouds [as] the most conspicuous examples.''
The Sm class was created when 
de Vaucouleurs (1959) described an extension of
``the spiral sequence into the irregular
types'' which he noted as ``SBd or SBm.''
This extension was motivated by the ``recognition of spiral structure in the
Magellanic Clouds and objects of similar type'' (de Vaucouleurs 1954).
The Sm class was later described by de Vaucouleurs
(1963) as ``the transition stage toward the Magellanic irregulars Im (whether
barred or not).''

The BCD class is not part of the RC3 morphological system, but rather
was imposed on galaxies after the identification of compact, high surface
brightness ``HII galaxies'' that are dominated by nebular
emission lines (Sargent \& Searle 1970).
Some
BCDs have properties similar to those of Im systems (Kunth 1985).
To select BCDs that are comparable to the
Im galaxies we
used the width
at 20\% intensity of the integrated HI profile W$_{20}$
to select systems that are
comparable in mass to Im systems (W$_{20}$$\leq$175 \kms).

The galaxies were chosen to be relatively nearby, and biased to systems
containing gas.
The sample is not complete, but it
spans a large range in galactic parameters and is representative.
The galaxy characteristics include a range in integrated
luminosity (M$_V$ of $-$9 to $-$19),
average surface brightness (20 to 27 mag/arcsec$^2$),
current star formation activity (0 to 1.3 M\solar\ yr$^{-1}$ kpc$^{-2}$),
and relative gas content (0.02 to 5 M\solar/L$_{B}$).
A more complete description of the sample and of its selection
is given by Hunter \& Elmegreen (2004) where we present the \ha\
imaging data.
Four galaxies observed in \ha\ are not included here because we
do not have broad-band imaging of them
(NGC 1705, NGC 2101, NGC 3109, F567-2), and two additional galaxies
(DDO 125 and Mrk 67)
have been dropped because they are interacting with companions.
On the other hand, two galaxies without \ha\ imaging are added
to the broad-band imaging sample (F473-V1 and F620-V3).

The distances to the galaxies in our sample and references
from which the distances are taken
are given in Table \ref{tab-sample}.
We used distances determined from
variable stars or the tip of the Red Giant Branch,
if they were available.
Other distances
were determined from the radial velocity relative to the Galactic
standard of rest V$_{GSR}$ (RC3) and a Hubble
constant of 65 \kms Mpc$^{-1}$. While this might not be the
most modern version of the Hubble constant and we do not correct
for Virgo-infall, we have retained the distances
that were used in the presentation of the \ha\ survey data
(Hunter \& Elmegreen 2004) and in several papers presenting
HI data (Simpson \et\ 2005a,b)
so that combining data from the various wavelengths is
straightforward.
Two galaxies---F473-V1 and F620-V3---have no distance determination.

For comparison to spiral galaxies we have used the sample compiled
by Kennicutt (1983). This sample of 74 galaxies spans the range of
morphologies from Sab to Sd. Kennicutt has measured \ha\ fluxes
for these galaxies. Other properties were obtained from RC3 and
Fisher \& Tully (1981). For the spirals, $\mu_{25}$, the average
B-band surface brightness within an isophote of 25 magnitudes in 1
arcsec$^2$ (discussed below in Fig. \ref{fig-histsb25}), has been
converted from the value in the RC3 to one that matches the
definition of surface area used here, i.e. the area of the
projected ellipse used for the denominator of the surface
brightness calculation in the RC3 has been replaced here with the
area of the circle having a radius equal to the semimajor axis.

Foreground reddening E(B$-$V)$_f$ was determined from Burstein \&
Heiles (1984) and values are given in Table \ref{tab-sample}. The
reddening law of Cardelli, Clayton, \& Mathis (1989) was adopted
with A$_V=3.1\times$E(B$-$V)$_t$, where the total E(B$-$V)$_t$ is
that due to foreground plus internal reddening. An
internal reddening correction of E(B$-$V)$_i$=0.05 was adopted for
all sample galaxies.

For the Kennicutt (1983) sample of 
spirals, an extinction correction was made to face-on orientation
using the formulation of Tully \et\ (1998) and the minor-to-major
axis ratios given by RC3. We then adopted a face-on internal E(B$-$V)$_i^0$ of
0.2 mag. Note that this is a change from our treatment of spirals
in Hunter \& Elmegreen (2004) where we adopted a single
internal reddening of 0.3 mag
after Kennicutt (1983) with no correction based on inclination. 
For 8 spirals the difference compared to our previous treatment is an increase
in E(B$-$V)$_t$ of 0.1--0.2 mag, and for 4 spirals the increase
is as high as 0.5 mag. But for the other 63 galaxies the change
is less than 0.1 mag, sometimes to a greater reddening correction
and sometimes to a smaller correction.

\section{The images} \label{sec-obs}

\subsection{UBV}

UBV images of our sample galaxies were obtained by one of us (DAH)
in 27 observing runs,
all but two of which took place 1997 to 2002.
The observations are listed in Table \ref{tab-obs}.
Most galaxies were observed in U, B, and V, but three have only
B and V and 6 have only V images.
One galaxy is shown in Figure \ref{fig-images} for illustration.
All of the data are available from http://www.lowell.edu/users/dah/images/,
and a few UBV color images constructed from the UBV can be seen at
http://www.lowell.edu/users/dah/ubv.html.

Most of the observations were made with a SITe
2048$\times$2048 CCD binned $2\times2$ on either the
Lowell Observatory 1.1 m Hall Telescope or the
Lowell Observatory 1.8 m Perkins Telescope.
Three galaxies---F565-V2, F563-V1, and F561-1---were observed
with a TI $800\times800$ CCD on loan from
the U. S. Naval Observatory, Flagstaff Station coupled with a
4:1 focal reducer.
In most cases several images were obtained of each galaxy in
each filter.
The telescope position was usually offset 20\arcsec\ between each
observation in order to average over flat-fielding defects.
The electronic pedestal was subtracted using the overscan strip,
and the images were flat-fielded using twilight sky flats.
The counts in the twilight sky flats were high enough that
they contribute much less than 1\% uncertainty to the galaxy images.
Landolt (1992) standard stars were used to calibrate the photometry.
For some galaxies observed on cloudy nights, frames of the galaxy
were taken on separate clear nights for purposes of calibration.
The multiple images of a galaxy in each filter were aligned and
averaged with an algorithm to
eliminate cosmic rays but preserve the photometric integrity
of the image.
The scale, seeing (FWHM of isolated star profiles), and
rms of the photometric calibration of the final images
are given in Table \ref{tab-obs}. The uncertainty in
the calibration is not included in the photometric uncertainty
of extracted quantities.

A few galaxies were kindly imaged for us by P. Massey
using the Kitt Peak National
Observatory (KPNO) 4 m Telescope and either a
2048$\times$2048 Tektronix CCD or the NOAO CCD Mosaic Imager, an array of 8
2048$\times$4096 SITe CCDs. In the case of the Mosaic Imager,
the galaxy was centered on and completely contained within one of the CCDs.
For the Mosaic Imager,
dome flats and twilight
sky flats were used to remove the pupil ghost as well as
determine the pixel-to-pixel variations.
The images were corrected for geometrical distortions and stacked
to produce a single final image in each filter.
(See Massey and collaborators' ``Local Group Survey:
Mosaic Reduction Notes'' at
http://www.lowell.edu/users/massey/lgsurvey/splog2.html
for details on how they have reduced Mosaic Imager data).
The Mosaic Imager data were calibrated
using field images and standard stars observed at the Lowell
Observatory 1.1 m Telescope.
P. Massey also obtained B and V images of IC 1613 for
us with the Cerro Tololo InterAmerican Observatory (CTIO)
Curits Schmidt 61 cm Telescope and a SITe $2048\times2048$ CCD.

In addition, Sextans A was observed by us with the CTIO 0.9 m
Telescope and these data are described by Hunter \& Plummer (1996).
IC 4662 and DDO 214 images were obtained with the CTIO 1.5 m and reported
by Hunter \et\ (2001b).
We also observed DDO 214 at Lowell Observatory, and we have
treated the two sets of observations of DDO 214 separately
for purposes of comparing photometry and other derived parameters.
However, the images obtained at Lowell Observatory go deeper and
are preferred for that reason, while the CTIO observations include U.
Images of NGC 6822 were taken by
C. Claver with the Prime Focus Direct Imager on the
CTIO 4 m Telescope and a Tektronix $2048\times2048$ CCD.
These data were calibrated with
images obtained with the Lowell Observatory 1.1 m Telescope.
Finally, V-band images of Haro 3 were obtained
as part of Service Observing on the KPNO 0.9 m Telescope
with a Tektronix $2048\times2048$ CCD.

Before performing surface photometry, we edited foreground stars
and background galaxies from the final UBV images, interpolating
across the edited region.
Usually, the V-band
image was done by hand, and then the cursor log file produced
from that was used to
remove objects from the other two filters. In that way, objects were
removed in the same manner from each filter.
One uncertainty in the galaxy surface photometry of the
outer regions is in distinguishing galactic from other
objects.

We then made a two-dimensional non-linear fit to the
background and subtracted it from the image to produce a
sky-subtracted final image for analysis (after Bell \et\ 2000a
and de Jong \& van der Kruit 1994).
Sky subtraction is an important source of uncertainty in our
surface photometry. Therefore, to gain confidence in our results, we
have explored the uncertainty in the sky.
First, we computed half of the
difference between the maximum and minimum in the fit to the sky over the
images (not just in the region of the galaxy but excluding a border
around the edges of the images). 
The average in this quantity is 1.2\% of the sky. This is the contribution
to the uncertainty that would result if we did not remove
the large-scale variation
at all, and obviously we have done better than that by fitting
and subtracting the background.
Second, we have taken the uncertainty of the sky from Poisson statistics
of the sky counts (formula given in \S3.3):
the square-root of the sky counts times the gain of the CCD. 
For each
of our images the Poisson statistics yields a greater uncertainty -- an
average of 2.2\% per pixel -- than those from the 
difference between the maximum and minimum in the sky fit.
Thus, we assume that our sky uncertainty is dominated by Poisson statistics,
especially after removal of the large-scale variations, and
Poisson statistics is the uncertainty that we use throughout
this paper.
Using this uncertainty
gives us confidence in the results that
stand out above the sky. This confidence will be important in our
discussion of outer disk profiles, where the galaxy intensity is
far below the sky intensity but well above the uncertainty of the
sky intensity. The average Poisson variation in the galaxy counts,
combined with the average uncertainty in the sky fit, suggests that the
average uncertainty in the photometric measurement of our galaxies
is of order 0.02--0.03 mag.

\subsection{JHK}

JHK images of 41 galaxies in our sample (26 Im, 12 BCD, and 3 Sm)
were obtained by one of us (DAH)
in 9 observing runs 1996 February to 1998 May.
Seventeen galaxies were observed in only J or, in the case
of DDO 35, only H. Twelve galaxies were observed in J and H,
and 12 in J, H, and K.
The observations are listed in Table \ref{tab-obs}.
One galaxy is shown in Figure \ref{fig-images} for illustration.
All of the data are available from http://www.lowell.edu/users/dah/images/.
These images were obtained using the Ohio State
Infrared Imager-Spectrograph (OSIRIS) on the Perkins 1.8 m Telescope.
The instrument consisted of four
mosaiced detectors to create a total array of 256$\times$256 pixels.
Observations were made with a pixel scale of 1.50\arcsec\ for
a total field of view of 6.4\arcmin.

Zero-second dark frames were subtracted from all of the data,
the image pixel values were corrected for non-linearity effects,
and pixel-to-pixel sensitivity variations were removed using
observations of a white screen in the dome.
Since OSIRIS consists of four detectors, separate
linearity corrections were determined for and applied to
each quadrant of an image. Non-linearity was of order 3\% at
levels of 15,000 counts.
We took images of the white screen in the dome both with and without
the screen illuminated, and subtracted the latter from the former in
order to remove stray background light before forming the flat-field
correction.

Observations of the galaxies consisted of several integrations of
approximately
5, 10, and 20 seconds for K, H, and J-bands, respectively.
For large galaxies that occupied most of the field of view,
the telescope was moved about 5\arcmin\
away for sky observations every few minutes or less.
Periodically the direction and length of
throw to sky were varied, and the center of the galaxy was moved
around in steps of 10\arcsec. The purpose of these steps was
to allow the pattern of stars in the sky observation and the locations
of hot pixels relative to the galaxy to change
so that they could be removed.
For small galaxies, the telescope was nodded between images to place
the galaxy alternately on one-half of the array and then the other,
allowing sky to be determined while also observing galaxy.
The object frames were sky-subtracted, aligned, and averaged.

``UKIRT'' standard stars
from the list compiled by S.\ Courteau were used
to calibrate the photometry, and so the photometry is on the
UKIRT system. The California Institute of Technology
(CIT) photometric system is 8\% bluer in J$-$H
and 4\% bluer in H$-$K compared to the UKIRT system.
Each star was observed 4 times with the star placed in each quadrant
of the detector array
in turn,
and sky for a specific observation was taken
from the observations with the star in the other quadrants.
We checked our photometry by comparing with the IR photometer
observations of NGC 4449 made by Hunter \& Gallagher
(1985a) with a 16\arcsec\ aperture. Our imaging data yielded
J and H magnitudes that were 0.24 magnitudes brighter than
those obtained by Hunter and Gallagher, but the J$-$H color
difference was zero.

We observed a few spiral galaxies in JHK for comparison.
These include NGC 3274 (SABd?), NGC 4027 (SB(s)dm),
NGC 4684 (S(r)B0$+$), NGC 4800 (SA(rs)b),
and UGC 2984 (SBdm$+$). We did not obtain UBV observations
for these systems. In addition, we observed the \HII\ region
NGC 5471 in M101 and the globular cluster NGC 6229 for
examples of a very young and a very old stellar population, respectively.

\subsection{Surface photometry}

Surface photometry was performed on all of the broad-band
and \ha\ images.
Because Im galaxies are lumpy, automatic ellipse fitting routines
seldom perform well on these galaxies.
Therefore, we determined the center of the
galaxy, position angle, and ellipticity from
a contour in the outer half of the V image which was
block-averaged by factors of a few
to increase the signal-to-noise.
The center is simply the geometrical center of
this isophote and the major axis is the longest bisector
that passes through the center
that, as much as possible, symmetrically divides the galaxy.
One exception was DDO 40 for which an inner contour was used instead
because the outer extended starlight did not seem to fairly represent
the bulk of the galaxy.

We then fixed these parameters and integrated in
ellipses that increase in semi-major axis length
in approximately 10\arcsec\ steps.
The center, position angle (P.A.), minor-to-major axis
ratio $b/a$, and ellipse step size are given in Table \ref{tab-obs}.
Bad pixels due to unsubtracted or poorly subtracted foreground stars,
satellite trails, or other flaws were masked and not used in the photometry.
The flux in each ellipse was scaled from the actual area used to the
expected area of the ellipse to account for such masked pixels.
In the cases of DDO 143, Haro 36, and Mrk 16,
much of the galaxy image was obliterated
by a very bright nearby star.

Since we wished to compare the surface brightness photometry between
these various images, we geometrically
transformed the UBJHK and \ha\ images to match the V image and used
the same elliptical integration parameters as were used for V.
The exception was that in a few cases \ha\ required more ellipses
(see especially, NGC 1569).

The surface photometry was fit with
$\mu = \mu_0 + 1.086 R/{\rm R}_D$ which represents an exponential disk
and R$_D$ is the scale-length of the disk.
The range of radii to fit was determined for each galaxy
from examination of a plot of $\mu_{V_0}$ or $\mu_{J_0}$ as a function of radius.
Deviations from a linear relation in the center or in the
outer parts or excursions due to a very large star-forming region
as in NGC 2366
were excluded from the fit. The fit was made with uniform weighting
of each radial data point.
In some galaxies $\mu_{V_0}$ is
best fit with two lines instead of one. In those cases, each fit was done
separately, and, again, the range of radii
to fit for each line was determined by examining the $\mu_{V_0}$(R)
plot.

The relative uncertainty in the surface photometry at each
annulus was determined assuming Poisson statistics, and is the sum
of the squares of the uncertainties in the galaxy counts in each
of the two ellipses that form the annulus and in the sky for each
ellipse, divided by the galaxy counts in the annulus, $f_a$:
\begin{equation}
\sigma_{\mu} = {1.086\over{f_a}} \sqrt{(f_2^{\prime} +
f_1^{\prime} + 2 s (A_1+A_2))/g},\end{equation}
where subscript
$1$ and $2$ refer to the first and second ellipses that bracket
the annulus in question, $f^{\prime}$ is the sky-subtracted
counts, $s$ is the average sky per pixel in the image, $A$ is the
area of the ellipse in pixels, and $g$ is the gain of the CCD. An
example of our surface photometry is given in Figures \ref{fig-sb}
and \ref{fig-colors}. NGC 2366 was chosen for this example because
it has a complete UBVJHK data set. Plots of all of the surface
photometry can be obtained from
http://www.lowell.edu/users/dah/sbplots.

Integrated photometry is given in Table \ref{tab-integrated}. The
integrated photometry is given at several different radii for each
galaxy, including the half-light radius R$_{1/2}^V$ and radii
appropriate to the JHK images. Usually, K does not go as deep as
H; H as deep as J; or J as deep as V. Therefore, photometry values
are reported at the maximum extents for reasonable uncertainties
in each of the JHK passbands. The largest radius listed is the
maximum extent of the V-band measurements.

In Table \ref{tab-structure} we report various structural parameters
measured from the surface photometry. This includes the exponential disk
fits to both V (first line) and J (second line), if available.
Some galaxy profiles were best fit with two lines, rather than
a single line. In these cases, R$_D$ and $\mu_0$ labeled as
``Primary'' are those derived from the fit to what we considered
to be the part of the profile that represents the bulk of the
underlying stellar disk. The second fit is labeled ``Secondary.''
R$_{Br}$ is the radius at which the two fits cross, and
the column labeled ``Sec?'' tells whether the secondary fit
is to the outer part of the profile (``O'') or
to the inner profile (``I'').
In the case of the BCDs, we have assumed that the inner part
of the disk is dominated by the intense star formation there
and so the outer galaxy
is a truer representation of the underlying
stellar disk (see, for example, Noeske \et\ 2003).
In the case of the Im and Sm galaxies that do not have
centrally concentrated and intense star formation, we take a
change in the outer disk, which usually occurs at low
surface brightness levels, to represent
a deviation from the basic underlying stellar disk,
as is seen in some spirals
(Pohlen \et\ 2002).

Double-exponential disks may contain important information
concerning star formation processes in galaxies.  A star formation
model that reproduces this profile is suggested in a companion
paper (Elmegreen \& Hunter 2005). Because breaks in surface
photometry profiles at low light levels could be caused by over or
under subtraction of sky, it is important to demonstrate that
these breaks are not artifacts of data handling. First, we note
that many of the breaks occur at relatively high surface
brightness levels where small errors in the determination of sky
do not have much effect.

Second, we have done ultra-deep imaging of three of our sample
of Im galaxies in order to examine the outer parts of the
disks in more detail (Hunter, Elmegreen, \& Anderson 2006). Two of
these galaxies show breaks in the surface photometry 
in our survey data presented here. The new data were taken with a
different instrument on a different telescope and processed
completely independently from the present survey. The
resulting new surface photometry shows the same breaks at the same
radii and surface brightnesses. The one galaxy in the deep-imaging
sample that did not show a break in the present surface
brightness profile also did not show a break in the new
data.

Third, we have verified that the double exponentials are present
in the data by making intensity cuts 11 pixels wide along
the major axes of seven of the systems. Differences between
the two sides of the galaxy are sometimes apparent in these cuts,
which is to be expected for irregular galaxies. However, the break
in the exponential profile is still clear on each side. This
result implies that the azimuthal-averages used for the
exponential fits in our full sample are not introducing false
breaks that make them look like double exponentials. Such false breaks
were suggested to be a possible source of smoothly tapering outer
disk structure by van der Kruit (1988), who noted that edge-on
galaxies usually have very sharp outer truncations. An example of
an intensity cut from our data is illustrated for DDO 48 in Figure
\ref{fig-cuts} along with cuts showing the original sky, the fit
to the sky, and the background after sky subtraction. We conclude
that the profile breaks and double-exponentials observed in our
sample of galaxies are real features of the disk structures.

In Table \ref{tab-structure} we report the radius at
which the reddening-corrected B-band surface photometry reaches 25
mag in one arcsec$^2$, R$_{25}$, in the projected disk.
We also give R$_H$, the Holmberg
radius, which was originally measured by Holmberg (1958)
at a photographic surface
brightness of 26.5 mag in 1 arcsec$^2$. Here we have used the conversion from
photographic magnitude to B-band magnitude listed in Table 11 of
de Vaucouleurs \et\ (1976; $=$RC2) $\mu_B = 26.5 + 0.22 - 0.149
(B-V)$, where (B$-$V) is averaged over those radii where the color
is well determined. For our galaxies this corresponds to a B-band surface
brightness of between 26.60 and 26.72 mag in 1 arcsec$^2$ with
an average of 26.64$\pm$0.002 mag in 1 arcsec$^2$. We also list
values of R$_{1/2}^V$, the radius containing half of the V-band
light of the galaxy.

In Table \ref{tab-stuff} we give $\mu_{25}$, the
average B-band surface brightness within the isophote defined by
R$_{25}$; $\mu_D^V$, the average V-band surface brightness
within the scale-length R$_D^V$; and $\mu_{2.5D}$, the average V-band
surface brightness within the radius 2.5R$_D^V$.
These three surface brightnesses are not corrected for projection.
They equal the galaxy flux divided by the
area of the circle with a radius equal to the semi-major axis
of the corresponding projected ellipse. This is in the
convention of Holmberg (1958). In \S\ref{sect:axes}, we introduce
another surface brightness, $\mu_{2.5R_D}$, which is the
flux divided by the area of the isophotal ellipse whose semi-major
axis equals $2.5R_D^V$.  These two surface brightness definitions
differ by the ratio of axes, with the latter value more
appropriate for comparison with the survey surface brightness limit
and models of apparent intensity.
Finally, in Table \ref{tab-stuff}
we also include the inclination of the galaxy $i$ derived from $b/a$
under the assumption that the intrinsic $(b/a)_0$ of an edge-on Im
system is 0.3 (Hodge \& Hitchcock 1966; van den Bergh 1988).

\subsection{Comparison with the literature}

Various of the galaxies in our sample have also been observed by
others
(Im galaxies: Aparicio \et\ 2000; Barazza \et\ 2001; Bremnes \et\
1998, 1999, 2000; Carignan \& Freeman 1988; Doublier \et\ 1999;
Karachentsev \et\ 1999; Lee \et\ 1999; Makarova 1999;
Patterson \& Thuan 1996; Thuan \et\ 1999; van Zee 2000;
BCD galaxies: Cair\'os \et\ 2001,2003; Doublier \et\ 2001;
Noeske \et\ 2003).
This gives us the opportunity to check our measurements against
those of others.

For 8 galaxies, our integrated U$-$B differ from those of others
on average by 0.5$\sigma$ (0.04 mag); B$-$V for 23 galaxies differ
by 1.2$\sigma$ (0.08 mag); and V by less than 3$\sigma$ (0.25 mag).
Here $\sigma$ is the quadratic sum of the uncertainties in our measurement
and those in the literature.
Results are similar for colors and magnitudes given by RC3.
Our R$_{25}$s differ on average from those given by RC3 by
0.4\arcmin, and R$_H$s differ from those measured by Holmberg (1958)
by 0.9\arcmin.
The average surface brightness within R$_{25}$ given by
RC3, and dereddened, differs from what we measure, converted to an elliptical area
normalization, by 0.5 mag arcsec$^{-2}$ on average for 74 galaxies.
From surface photometry,
$b/a$ is on average 4\% different from those used by others
for 24 galaxies, and the position angle differs by up to 40\arcdeg\
for 33 galaxies.
For 21 and 22 systems, our $\mu_0^V$ and R$_D^V$ differ from others by
4$\sigma$ on average (0.9 mag and 19\arcsec, respectively).

Another useful comparison comes from the galaxy DDO 214
for which we obtained two sets of images, one with
the Lowell Observatory 1.1 m Telescope and one with the
CTIO 1.5 m Telescope. We treated these two data sets independently,
although the Lowell Observatory data go deeper.
The V magnitudes integrated over the galaxy differ
by 0.06 mag, and B$-$V by 0.16 mag.
The inferred $b/a$ ratios differ by 0.04, or 5\%,
and the position angles by 4.5\arcdeg.
From surface photometry, the $\mu_0^V$ differ by 0.18 mag 
arcsec$^{-2}$ (1.6$\sigma$) and R$_D^V$ by
3.6\arcsec\ (1.2$\sigma$, 11\%).
The radii R$_{25}$ and R$_H$ differ by 1\arcsec\ ($<$1\%, $<1\sigma$), and
R$_{1/2}^V$ by 2.6\arcsec\ (6\%, $<1\sigma$).
The $\mu_{25}$ differ by 0.11 mag arcsec$^{-2}$,
$\mu_D^V$ by 0.07 mag arcsec$^{-2}$,
and $\mu_{2.5D}$ by 0.2 mag arcsec$^{-2}$.
Thus, except for B$-$V, the comparisons are reasonable.
They suggest that an estimate of the total uncertainty
in photometric quantities, including calibration
(when not limited by Poisson statistics),
is of order 0.03 mag, the same as we concluded in \S3.1.

\section{Results} \label{sec-results}

\subsection{General Trends}

Histograms showing the number distribution of the sample
in M$_{V_0}$, the average B-band surface brightness within R$_{25}$, $\mu_{25}$,
and the average V-band surface brightness within R$_D^V$, $\mu_D^V$, are
shown in Figures \ref{fig-histmv}, \ref{fig-histsb25},
and \ref{fig-histsbdv}.
Figure \ref{fig-histhalfr} shows the number distribution
of the sample in R$_{1/2}^V$, the radius that contains
half of the light of the galaxy in the V-band.

\subsection{General morphology}

\subsubsection{Peculiarities}

Irregular galaxies are, of course, morphologically irregular.
Even so, there is an orderedness to most Im systems.
However, some irregular galaxies exhibit peculiar features that
go beyond the norm for the class, and these are identified in
Table \ref{tab-stuff}.
Several of the galaxies are curved or have crescent shapes: DDO 215 and
F533-1.
There are a few that have elongated structures extending from or
curving around
part of the galaxy
(DDO 25, DDO 169, F620-V3, NGC 2552, UGC 5716, UGC 8276)
that look like spiral-arm fragments or tidal tails.
Others have broad, asymmetrical extensions
(DDO 27, DDO 35, DDO 40, DDO 63, DDO 99, DDO 214, F651-2)
or peculiar twists (DDO 48).
One galaxy, DDO 165, has a sharp, curved southern edge.
Some galaxies are simply messy with strong central regions and
lower surface brightness scatter
(DDO 9, DDO 34, DDO 68, UGC 199, UGC 8055, UGC 11820).
In all, 18 of the 94 Im galaxies (19\%) have noticeable
large-scale morphological peculiarities, as do 5 of the 18 (28\%)
Sm systems. Oddly, none of the 24 BCD systems show these kinds of
peculiarities even though the unusual intensity and pattern of
star formation of the class suggest that these galaxies have incurred
interactions. The sample with morphological peculiarities
has the same distribution of current star formation rates as the
rest of the sample, but tends to be bluer.

We have compared the position angles of the morphological major axis
of the \HI\ and of the optical components of those survey galaxies
with \HI-line interferometric data in the literature.
We eliminated galaxies
with $b/a\geq0.9$ since the optical
position angle is poorly determined when the image gets round. We have
also eliminated galaxies with no clear rotation since the \HI\ kinematical
axis is then poorly determined. That left 31 Im and 6 Sm galaxies (no
BCDs).
Of these the majority have position angle differences that
are less than 30\arcdeg; seven galaxies have differences
of 40--80\arcdeg. Of these 7 galaxies, 3 are barred. Since
bars can rotate with respect to the rest of the disk,
a misalignment between the major axis of the bar and
the outer disk is not unusual.
The misalignment in the remaining 4 galaxies (DDO 26, DDO 53,
DDO 86, DDO 168) may indicate a past disturbance. These galaxies,
13\% of the Ims with \HI\ interferometric data,
are not marked as having morphological peculiarities, but
the \HI\ kinematics are also peculiar in two of them
(DDO 26---Hunter \& Wilcots 2002, DDO 168---Broeils 1992).
These galaxies, combined with the fraction that show unusual morphological
structures, mean that as many as one-third show some abnormality.

The most common phenomena that would account for
these distortions to galaxy disks are interactions with other
galaxies or extragalactic \HI\ clouds.  This is in spite of
the fact that our sample is biased against obviously interacting
systems. Dwarfs are also much more
susceptible than giant spirals to disruption by the pressures of
star formation. This follows from the low surface densities and
interstellar pressures in dwarf galaxies, the lower levels of dust
absorption at low metallicity, and the brighter uv fluxes from
metal-poor massive stars.  Interstellar turbulence at typical
velocities should also be more disruptive in dwarfs than in spiral
galaxies because the rotation speeds of the dwarfs are lower.
When the turbulent speed is a large fraction of the rotation
speed, the epicyclic radius becomes a large fraction of the disk
scale. Then turbulent excursions that produce shells, holes, and
cloud complexes are large compared to the disk scale-length,
resulting in a more irregular overall appearance.

\subsubsection{Bars}

Some fraction of disk galaxies are barred, and irregular galaxies
are no exception. However, identifying bar structures in irregular
systems is harder than in spirals because the symmetry provided by
spiral arms is missing. Furthermore, bars in Im galaxies can be
comparable in size to the optical galaxy
(NGC 1156: Hunter \et\ 2002; NGC 2366: Hunter, Elmegreen, \& van Woerden
2001a; NGC 4449: Hunter, van Woerden, \& Gallagher 1999),
so a large-scale boxy appearance due to a bar
could be hard to distinguish from inclination effects.
The contour plots that were used to determine the surface photometry
parameters were also examined for the signs of a bar structure.
A boxy appearance turning to rounder isophotes in the outer parts
and/or a twisting of isophotes from the inner galaxy to the outer galaxy
were considered strong signs of a bar. Bar structures were
often apparent in color-color ratio images as well, since many
of the bars are blue and lined with \HII\ regions.

If a galaxy appears to be barred, the bar length, minor-to-major
axis length ratio, and change in position angle between the bar
and the outer galaxy, if any, are given in Table \ref{tab-stuff}.
Characteristics of the bar were measured from a contour plot of the V-band
image that was block-averaged by factors of a few to increase
the signal-to-noise. The edge of the bar was taken to be the contour
at which the shape ceased to be boxy and became more spherical and/or
where the position angle changed. These changes were generally obvious
but rarely sharp. The ambiguity in the end of the bar is quantified
in the uncertainty in the bar's length.

In all, 22 of our 94 Im galaxies (23\%) show evidence for a bar
structure while 3 of 24 BCDs (12\%) and 9 of 18 (50\%) Sms do.
Given the sample sizes, the percentage of BCDs and Sms with
bars is not more than one sigma different from that of the Ims.
The semi-major axis of the bar R$_{Bar}$ relative to the disk
scale-length R$_D^V$ is shown in Figure \ref{fig-rbrd}.
R$_{Bar}$ is most often 1.5--2R$_D^V$ for Im galaxies;
the range is 0.85R$_D^V$ to 3.5R$_D^V$ and 7 of the 22 barred
Im galaxies have R$_{Bar}$ greater than 2R$_D^V$. Thus, the bars in Im
galaxies are relatively large, and in some cases occupy most of the
optical galaxy. The bars in the Sm sample are smaller with a
typical length of 1--1.5R$_D^V$ and only one galaxy has a bar
greater than 2R$_D^V$. Bar minor-to-major axis ratios $(b/a)_B$ range
from 0.33 to 0.80 for Im galaxies with a typical value being 0.6--0.7.
Bars in our Sm galaxies have similar shapes.

We have made profile cuts through the major and minor axes of
the bars, centered on the optical center of the bar determined
from the bar's outer isophotes in V. These profiles are
illustrated in Figure \ref{fig-barcuts}.  In some cases, the major and minor
axes profiles are similar to each other, meaning that the bars are
relatively thick. Most of the major axis profiles are exponential,
as in late type galaxies in general (Elmegreen \& Elmegreen 1985), but a few
are flat, such as DDO 35, DDO 154 and DDO 133. There are no other
obvious difference between these two cases.

Because of the spectacular offset bar in the nearby Im galaxy LMC,
it is often assumed that bars in Im galaxies are generally offset from the
galaxy centers. In our sample
of barred galaxies, we found that the optical centers of
the bars in 68\% of the Im galaxies,
100\% of the BCDs, and 67\% of the Sms lie within 500 pc
of the center of the galaxy defined by the outer V-band isophotes
in the plane of the sky.
Only 14\% of the Ims (3 galaxies) and 22\% of the Sms
(2 galaxies) have bars whose centers lay more than 1 kpc
from the optical center of the galaxy.
The majority of Im galaxies with measurable offsets
(57\%, 8 galaxies) have offsets that are less than half of a disk
scale-length. Only one has an offset that is greater than R$_D^V$.
The 4 Sm galaxies with measurable offsets lie between 0.4R$_D^V$
and 0.6R$_D^V$.
We conclude that the offset feature of bars is not a general phenomenon
of late-type galaxies.

Bars in Im galaxies appear to trigger a substantial amount of star
formation. Figure \ref{fig-barsIm} shows histograms of the integrated star
formation rates, the average surface brightnesses
inside R$_D^V$, and the ratios of the radial extents of the \HII\
regions to R$_D^V$ for the barred and unbarred samples.
The medians for these three
quantities, respectively, are $-2.2$ \sfrunit\ (the logarithm of the integrated
star formation
rate normalized to the area within R$_D^V$),
$23.2$ V-magnitude arcsec$^{-2}$,
and $3.0$ for barred Ims, and $-2.7$ \sfrunit,
$24.1$ V-magnitude arcsec$^{-2}$, and
$2.2$ for non-barred Ims. Bars apparently increase the
areal star formation rates and surface brightnesses by a factor of
2 to 3, and they increase the radial extents of the \HII\ regions by
a factor of $\sim1.4$. This increase may simply reflect the
increased density of gas in a bar compared to a non-barred central
region, although gas shocking in a bar flow or gas pileup at the
end of a bar could be additional triggers (Elmegreen \& Elmegreen 1980).
Slight shifts like these in star formation properties are also
seen for BCD galaxies, but the statistics are too poor to make a
conclusion. There is no obvious trend for the Sms.

\subsubsection{Minor-to-major axis ratios}
\label{sect:axes}

Past studies of the distributions of projected optical
minor-to-major axis ratios
$b/a$ have derived the intrinsic shape of Im galaxies
under the assumptions that there is one intrinsic shape and
that galaxies are oriented at random on the sky. However, different
studies have come to quite different conclusions.
Hodge \& Hitchcock (1966) and van den Bergh (1988) found
that Im galaxies are modestly thick disks with an
intrinsic ratio $(b/a)_0$ of
0.3--0.4 rather than the 0.2 value generally adopted for spirals.
Staveley-Smith, Davies, \& Kinman (1992)
suggested from gas kinematics that the disk is thicker than this,
with $(b/a)_0\sim0.6$.
On the other hand, Binggeli \& Popescu (1995) and Sung \et\ (1998)
concluded that Im galaxies are triaxial in shape, similar to
dwarf ellipticals.

The distribution of projected $b/a$ for our sample of galaxies is
shown in Figure \ref{fig-ba}.  The statistical uncertainties are
relatively low only for Im galaxies, which show a peak at
$b/a=0.5$ to 0.6. Although intrinsically flat, some edge-on Im galaxies
might be mis-identified as type Sm or even in some cases Sd. In any case,
the decrease at low $b/a$ in Figure \ref{fig-ba} is consistent with an
intrinsic thickness equal to $\sim0.3-0.4$ times the major axis
length (Hodge \& Hitchcock 1966; van den Bergh 1988). The decrease
at large $b/a$ could be caused by an intrinsically triaxial shape.
The random projection of circular disks would give a flat
distribution of $b/a$ up to $b/a=1$, which is not observed.

Figure \ref{fig-wl2} shows a model for the distribution of axial
ratios that fits the observations for Im galaxies.  The model
calculates the distribution of projected $b/a$ for random
orientations of triaxial ellipses that have intrinsic ratios of
width (W) to length (L) uniformly distributed in the range 0.7 to
1, and intrinsic ratios of thickness (Z) to length uniformly
distributed in the range 0.29 - 0.67 (see discussion of triaxial
models in Elmegreen et al. 2005). If this is the explanation for
the distribution of apparent axial ratios in Im galaxies, then
they are triaxial with an average ratio of around $L:W:Z =
1:0.85:0.48$. This is similar to what Sung \et\ (1998) found---$L:W:Z =1:0.7:0.5$
(see also Binggeli \& Popescu 1995).

However,
Figures \ref{fig-sblongrd_bamv} and \ref{fig-babmva_longrd} suggest the
situation is slightly more complicated than this.  Here
we plot $\mu_{2.5R_D}$,
integrated M$_{V_0}$, 2.5R$_D^V$, and (B$-$V)$_0$, all as functions of $b/a$.
Here $\mu_{2.5R_D}$ is the average V-band surface brightness
within $2.5R_D^V$, normalized to the area in the ellipse that
integrates the light at 2.5R$_D^V$, rather than the circular area in the plane
of the galaxy used for the values given in Table \ref{tab-stuff};
$\mu_{2.5R_D}$ is smaller (brighter) than $\mu_{2.5D}$ listed in
Table \ref{tab-stuff} by $2.5\log\left(a/b\right)$.
The three galaxy types are
plotted with different symbols and colors, and the lines are
averages for each type in intervals of 0.1 in $b/a$.

The top part of Figure \ref{fig-sblongrd_bamv} indicates that the
$\mu_{2.5R_D}$ values for Im galaxies at the peak of the $b/a$
distribution are clustered near the limit of detection, which is
at the lower part of the panel. The two figures further show
that the sizes and luminosities are smallest at the peak
of the $b/a$ distribution. Outside of these regions in the
figures, the distribution of $b/a$ values is more uniform. This
suggests that the smallest and faintest Im galaxies could be
triaxial, but the larger and brighter Ims are more disk-like.

An alternative possibility for this distribution is that we have
observed only the edge-on examples of the Ims that have very low
intrinsic surface brightness.  An edge-on projection of a disk has
a higher apparent surface brightness than the face-on
projection, in proportion to the ratio of optical path lengths.
Thus the extremely faint Im's could be visible only when viewed
edge-on (see Elmegreen et al. 2005).

The bottom part of Figure \ref{fig-sblongrd_bamv} indicates that all
galaxy types in our sample brighten slightly as $b/a$ increases,
and correlation coefficents are consistent with this.
The increase follows from the near-constant values of surface brightness
and size as a function of $b/a$ for each
type. Larger $b/a$ then corresponds to larger projected area and
more exposed galaxy.  Such a correlation is expected to appear
whether or not we select against the face-on versions of the
lowest surface brightness galaxies.  It would not appear if the
galaxies were optically thin, but the largest
increase is for the BCDs, which probably have the most opacity.

The bottom part of Figure \ref{fig-babmva_longrd} suggests that (B$-$V)$_0$
for BCD galaxies reddens with increasing $b/a$.  This color does
not change noticeably for the other types.  The origin of this
seems to follow from the radial distribution of colors for BCDs
(see Figure \ref{fig-shallowouter_bcd} below) if the inner starburst parts of
these galaxies are more spherical than the outer old red disk. Then
the change from edge-on to face-on will brighten up the red outer
disk but not brighten the inner blue starburst as much.  The result
is an overall redder color along with the total brightening for
more face-on cases.

\subsection{Colors}

\subsubsection{Integrated photometry}

The integrated colors are shown in color-color plots:
UBV in Figure \ref{fig-ubv}, JHK in Figure \ref{fig-jhk},
and VJH in Figure \ref{fig-vjh}.
The irregular galaxies are bluer in UBV than spirals, as is well known
(de Vaucouleurs, de Vaucouleurs, \& Buta 1983).
There is, however, a large range---0.5 mag---in the UBV colors within
the sample.
In J$-$H, the colors fall between those of an \HII\ region, which represents
a very young stellar population, and those of a globular cluster,
which represents a very old stellar population. The spirals that
were observed for comparison lie near the globular cluster.
The H$-$K colors
of these objects are not well separated.

\subsubsection{Gradients}
\label{sect:gradients}

Most Im galaxies are remarkably uniform in color, showing
no change with radius (see, for example, Bergvall \et\ 1999, van Zee 2000).
However, this is not universally true of the class.
In Makarova \et's (2002) study of 16 nearby dwarfs, about
half showed some gradient, becoming redder in the outer parts.
Similar results were found by Parodi, Barazza, \& Binggeli (2002).
Of the 94 Im galaxies in our sample,
44 (47\%) show a radial gradient in at least
one color. Surface photometry with a gradient that
could be equally well fit with a constant color within the
uncertainties is not counted. Also, a variation from
a constant color of only 1--2 annuli was not counted as a gradient since
these are probably due to individual star-forming complexes.
Gradients are also seen in 13 of 24 BCDs (54\%) and 14 of 18
(78\%) of our Sms.
For these galaxies, the color gradients are given in Table
\ref{tab-colgrad}.
The gradients, when present, however, tend to be small.
The minimum gradients that we detected were 0.02 mag kpc$^{-1}$
in B$-$V and U$-$B, 0.04 mag kpc$^{-1}$ in J$-$H and H$-$K,
and 0.09 mag kpc$^{-1}$ in V$-$J.
The typical gradient, when present, in B$-$V
is 0.1 mag kpc$^{-1}$ and in V$-$J is 0.3 mag kpc$^{-1}$.

For comparison, de Jong (1996c) examined the BVRIK color gradients in
86 face-on spiral galaxies. All galaxies became bluer with radius,
and all colors were strongly correlated.
Gadotti (1999) has also examined color gradients
in Sbc spirals. He found that 25\% had
no gradients, but most of these are barred and he suggests
that the stellar bar potential induces radial mass flows
to the center, serving to homogenize the stellar population.
Of the rest, 65\% have gradients that become redder to the center
and 10\% become bluer to the center.
Bell \& de Jong (2000) and MacArthur \et's (2004) analysis of
large samples of spiral galaxies showed
``significant'' color gradients in all of the galaxies with
most being redder in the central parts.

One striking feature of those galaxies in our survey showing
measurable gradients
is the great variety in the profiles.
Among the Im galaxies with color gradients,
67\% (24 galaxies) with a gradient in B$-$V 
become redder with radius and 56\% (20 galaxies) with
a gradient in U$-$B become redder.
By contrast most (10 of 15 gradients, 8 of 9 galaxies)
become bluer with radius
in J$-$H, H$-$K, or V$-$J.
About 26\% of the UBV profiles (in 15 galaxies) are complex,
illustrated with NGC 2552 in Figure
\ref{fig-colorsvar}, becoming redder or bluer or
staying flat at different radii within a galaxy.
Among the BCDs with color gradients, 75\% become redder with radius
in B$-$V and 91\% in U$-$B. All (6 gradients in 5 galaxies)
become redder in the VJH colors as
well. Of these systems with B$-$V gradients
42\% have complex profiles, and one
also has a complex U$-$B profile.
Of the Sm systems, 87\% of the color gradients in the UBVJH colors
become bluer with radius.
Six galaxies (40\%, 8 gradients) have complex profiles.
MacArthur \et\ (2004) saw similar behavior among their spiral
galaxy sample: The color gradients in the inner part and outer
part sometimes had different slopes, and sometimes changed sign.

Bell \& de Jong (2000) and MacArthur \et\ (2004) have examined
the relation between age gradients in spirals, derived from color
gradients, and various galactic parameters. They find a
correlation between age gradients when the radius is
normalized to the disk scale-length and the luminosity and size
of the galaxy. Bell and de Jong also find a correlation with
the central surface brightness which MacArthur \et\ do not see
when the gradient is normalized to the scale-length of the disk.
In Figure \ref{fig-colorprofs} we plot the B$-$V color gradient,
normalized to R$_D^V$, against these three galactic parameters.
The black symbols are the second parts of two-part gradients
(see Table \ref{tab-colgrad}). Most of the Im galaxies, covering
the entire range of observed M$_V$, have zero
gradients.
However, for those galaxies in our sample that do exhibit gradients,
we see that the color gradient correlates with M$_V$ only in
the sense that the Sm galaxies are both systematically more luminous
and also systematically have negative color gradients (redder in the
central regions). Similarly, there is a correlation of color
gradient with R$_D^V$ only in the sense that the Sm systems
extend to bigger systems compared to BCDs and Im galaxies and
dominate the negative side of the plot. However, it is plausible
that the spiral galaxies, which are more luminous and larger than
our sample galaxies, would extend the Sm part of parameter space
on our plots, placing our sample as part of the correlations seen
by these researchers.
Like, MacArthur
\et\ we see no correlation with $\mu_0^V$.

Another striking feature of some of the survey galaxies is that
color gradients are not correlated among the different
pass-bands as might be expected if different colors emphasize
different stellar populations.
An example of this is shown for NGC 2552
in Figure \ref{fig-colorsvar}.
In addition, three of the BCDs have B$-$V
and U$-$B profiles with opposite gradients.
We have not measured the JHK colors
as far out in radius as the UBV colors, so we do not know what the
JHK are doing in the outer parts, where one might see changes
in the optical colors.
However, among the galaxies with J, H, or K measurements in addition
to UBV, at least one JHK color, including V$-$J, does not show the
same sort of general trends or variations as
the optical colors in 12 of the Im galaxies,
while in another 14 the colors are correlated. 
Among the BCDs, 1 shows different passbands doing different things,
and in 11 systems the colors track each other.
Among the Sms, 2 galaxies show different trends in different colors
and in 1 they follow the same trend.
In most spiral galaxies, on the other hand, the colors
are strongly correlated (de Jong 1996c). However
the color-color diagrams in Bell \& de Jong (2000) and MacArthur \et\ (2004)
show kinks in some of the radial profiles of spirals as well.

Color gradients are the result of a mixture of stellar age
gradients, metallicity gradients, and extinction gradients. The
generally blue centers of the BCDs are probably the result of
recent star formation following some major inflow of gas (Hunter
\& Elmegreen 2004). The mixed color gradients in the other types
may reflect a patchy history of star formation.
Extinction gradients
should cause all of the disk colors to vary in the same way with
radius.  Figure \ref{fig-bruzual} shows B$-$V and V$-$J colors
from Bruzual \& Charlot (2003) models using the Padova1994
evolutionary tracks and the Chabrier (2003) stellar initial mass
function (IMF). The five
different curves are for different metallicities, as indicated by
the colored lines. On the left are the galaxy colors for a single
burst stellar population that has aged for the time indicated on
the abscissa. On the right are models with star formation rates
decaying exponentially, as $\exp\left(-t/{\rm decay time}\right)$, starting 10
Gyr ago (at $t=0$) and continuing until today (at $t=10^{10}$ yr).
When the decay time is small, the stars are all old and the colors
resemble the single stellar population models for large ages. When
the decay time is large, the models are like continuous star
formation. Starburst models have colors indicated by the low age
asymptote in the left-hand panels.  The figure has the usual
result that both increasing age and decreasing metallicity cause a
population of stars to redden. However, there are cases where
B$-$V can get redder while V$-$J gets bluer, as observed for
3 of our Im galaxies.  This may happen, for example, when
the inner region has a dominant age of $\sim10^7$ yr and a moderately
high metallicity (e.g. $Z=0.008$) while the outer region has a
dominant age of $10^8-10^9$ yr and a lower metallicity.
Thus, the changes in color gradients within a galaxy probably reflect
the large-scale bubbling of star formation activity around the
galaxy over time, and the uncorrelated behavior of colors
reflects how recently star formation has occurred in a given annulus.

\subsubsection{Two-dimensional color-color images}

Azimuthally-averaged color profiles are not always an adequate indicator
of color structure in the galaxy.
For example, color profiles are uniform with radius in NGC 2366 and NGC 4449.
Yet, Hunter \et\ (1999, 2001a) found that both of these barred
systems have a bright, blue ridge that crosses from one corner
of the rectangular inner bar to the opposite corner.
Figure \ref{fig-cross} illustrates this with color-ratio images.
These are the only two Im galaxies that we have found that
show this kind of color structure, but they do provide a caution
that motivated us to carefully examine two-dimensional color-color ratio
maps of each galaxy in our sample.
(All of the color ratio images are available from
http://www.lowell.edu/users/dah/colormaps/).

For most of our galaxies the two-dimensional color ratio images
look as expected: The azimuthally-averaged colors are flat
with radius and
the two-dimensional ratio images are uniform. The optical
UBV ratio images are punctuated with
blue spots that correspond to \HII\ regions while the JHK ratio images
do not show structure.

Some variations from this picture are the natural consequence of
variations in the evolutionary status of individual star-forming regions.
In some galaxies, especially the BCDs,
some \HII\ regions are associated with red knots
in B/V or U/B (A1004$+$10, DDO 27, F533-1, Haro 4, Haro 8,
Haro 20, Haro 29, Haro 38, VIIZw 403).
Presumably the red knots are dominated by red supergiants,
evolved massive stars.
There are also blue knots
(DDO 50, DDO 120, DDO 167, DDO 187, DDO 216, F564-V3, M81dwA,
NGC 1569, NGC 2552, NGC 3413, NGC 3738, WLM)
and red knots
(CVnIdwA, DDO 25, DDO 215, F721-V2, Mrk 178, NGC 4214) that
do not correspond to \HII\ regions, and these probably represent regions
that are old enough for the \HII\ region to have dissipated.
Interestingly, there are also blue rings with red holes
(DDO 33, DDO 155, Haro 3, Mrk 5, Mrk 408) that
are sometimes associated with an \ha\ hole or ring
of \HII\ regions (Mrk 16, Mrk 757, VIIZw 403).
These are probably instances of star-induced star formation,
with the blue ring representing the second generation.

On a larger scale, others have noted that some irregular galaxies
have extended red stellar populations while the blue stars are
more concentrated to the center of the galaxy
(WLM: Minniti \& Zijlstra 1996;
DDO 210: Lee \et\ 1999;
NGC 3109: Minniti, Zijlstra, \& Alonso 1999;
DDO 190: Aparicio \& Tikhonov 2000;
DDO 187: Aparicio, Tikhonov, \& Karachentsev 2000;
Makarova \et\ 2002; Hidalgo, Mar\'in-Franch, \& Aparicio 2003b).
This is especially true of BCDs in which the current star
formation is often intense and
centrally located
(Kunth 1987; Papaderos \et\ 1996; Thuan, Izotov, \& Foltz 1999;
Cair\'os \et\ 2003; Noeske \et\ 2003).

In our sample, we also see that some
of the galaxies have blue centers and relatively red
($\Delta$(B$-$V)$\sim0.5$) or somewhat redder
($\Delta$(B$-$V)$\sim$0.1--0.2) outer parts.
In the cases
of A1004$+$10, DDO 70, DDO 120, DDO 168, DDO 216, F565-V1,
Haro 23, Haro 38, Mrk 32, Mrk 408, Mrk 757,
NGC 1569, NGC 3738, NGC 3952, NGC 4163, NGC 6789, WLM, and VIIZw 403,
this color structure is clear in the
azimuthally-averaged surface photometry.
But, for Haro 36 and NGC 3413 the surface photometry does not show this
because the blue structures are elongated along the major axis.
However, other galaxies have relatively red ($\Delta$(B$-$V)$\sim$0.1--0.3)
centers and bluer outer parts
(DDO 9, DDO 24, DDO 48, DDO 50, DDO 88, DDO 105, DDO 122, DDO 173,
DDO 180, DDO 204, DDO 214, F563-V2, F651-2, NGC 2552, NGC 3510,
UGC 5716).

Some galaxies show large-scale blue or red regions that are
not aligned with the center
and some of these regions occupy nearly half of the galaxy
(DDO 155, DDO 165, DDO 171, Haro 3, UGC 5209).
This was also seen for the BCD Tol 3 by Noeske \et\ (2003).
Sometimes the \HII\ regions are not located in the part of the
galaxy that is blue in B$-$V
(F620-V3, Haro 14, HS 0822$+$3542).
The large-scale patchiness in colors most likely
indicates bulk variations in the stellar populations, and are
clear indications of the manner in which star formation
can bubble around the galaxy on scales of kiloparsecs
(Hunter \& Gallagher 1986).

DDO 165 and NGC 3413 merit a special note for their peculiar
color structure. DDO 165 has a very sharp, curved southern edge
and this edge is bluer than the rest of the galaxy.
NGC 3413 has a narrow, high surface brightness blue ridge down the middle
of a redder, broader disk. The blue ridge extends the length
of the galaxy, but \HII\ regions are found only in the
center.

Galaxies with some type of peculiar color structure are flagged in
Table \ref{tab-stuff}.
Of the Im group 23 (24\%) have some peculiar color structure,
while 11 (46\%) of the BCDs and 10 (56\%) of the Sms do.

\subsubsection{Relationship between colors and gas surface density}

We have examined the relationship between changes in the
azimuthally-averaged color profiles
and variations in the gas surface density. The motivation here is
that the gas is the fuel for star formation, and one might
expect that regions with a greater reservoir of fuel would
more readily form stars and that this would be reflected in
the bulk stellar populations as revealed in the optical colors.
The \HI\ profiles that are available in the literature
for galaxies in our survey
are shown in the figure
discussed in \S 4.4.3 and the references are given in that figure
caption.

We consider the ratio of the observed gas surface
density \siggas\ to the critical gas density \sigcrit\ derived
from the gravitational instability models for differentially-rotating disks
(Safronov 1960, Toomre 1964, Quirk 1972).
In these models there is a critical column density \sigcrit\
above which the disk is unstable to ringlike perturbations
and can easily form star-forming clouds, and below which the disk
is stable and less likely to form clouds. The rotating gas disk model was applied
by Kennicutt (1989) to spiral galaxies to explain the apparent
drop off in star formation in the outer regions (see more detailed models
of this drop-off in Elmegreen \& Hunter 2005).
Hunter \& Plummer (1996), Meurer \et\ (1996), van Zee \et\ (1997a),
and Hunter, Elmegreen, \& Baker (1998) have applied this model
to Im galaxies and found that it fails
to predict where star formation occurs in Im galaxies.
However, we now
take a fresh look at the relative \siggas/\sigcrit\ in the context of color profiles
within galaxies.
We use the \sigcrit\ and \siggas/\sigcrit\
derived by Hunter, Elmegreen, \& Baker (1998)
for 12 of the galaxies in our survey.
\sigcrit\ $= A c \kappa/\pi G$, where $\kappa$ is the
epicyclic frequency, $c$ is the velocity dispersion of the gas, and $A$
is a constant determined by Kennicutt to be of order 0.7.
The $\kappa$ were derived from the \HI\ rotation curves whose references
are given in our earlier study. There were observed velocity dispersions
available for 4 galaxies in the range from
6.8 to 9.5 km s$^{-1}$; the rest were assumed to be 9 \kms.

We find no clear connection between \siggas/\sigcrit\ and azimuthally-averaged
colors.
In IC 1613 the outer part of the galaxy does begin to become
redder in B$-$V and U$-$B just where \siggas/\sigcrit\ and
$\Sigma_{H\alpha}$ begin to drop in value.
A lower value of \siggas/\sigcrit\ should mean that it is harder
to form stars, and hence the redder stellar population would
be consistent with this.
On the other hand,
in DDO 155 there is a small gradient to redder colors in the outer
parts where there is a gentle trend to {\it higher} \siggas/\sigcrit.
In NGC 2366 and DDO 154,
the colors and \siggas/\sigcrit\ are fairly constant with radius.
On the other hand, in DDO 50 there is a large abrupt change in \siggas/\sigcrit\
at a radius of 2.75\arcmin\
accompanied by only a minor, smooth trend in B$-V$ and constant
U$-$B. Similarly, there is a rather large change in \siggas/\sigcrit\
in DDO 105, DDO 168, and F563-V2
where there is only a very small ($<$0.2 mag) change in colors.

We also examined the possibility of a correlation between the
colors and the simple gas column density \siggas,
but again we do not see a consistent
pattern. However, \siggas\ is relatively flat within the confines
of the optical galaxy in 8 of 9 galaxies (see \S 4.4.3 and the figure
discussed there); in these systems
\siggas\ drops by a factor of 1.5---3.8 over the region that
colors are measured. DDO 168 is the
exception in that \siggas\ drops by a factor of 9.8 over
the radius of the optical galaxy,
but the optical colors are, nevertheless, relatively constant
with radius.

\subsection{Exponential disks}

\subsubsection{Disk scale-lengths and surface brightnesses}

Hodge (1971) pointed out that Im galaxies possess
exponential disks, and all of the surface brightness profiles of
the galaxies in our sample are
well fit, at least in part, with an exponential.
Histograms showing the number distributions of the V-band scale length
R$_D^V$ and central surface brightness
$\mu_0^V$ from the fits to surface brightness profiles are
shown in Figures \ref{fig-histrdv}
and \ref{fig-histmu0v} (see also van der Kruit 1987).
Figure \ref{fig-histrdv} can be compared to Figure 15 of
Swaters \& Balcells (2002) which presents R-band disk scale-lengths
for a large sample of Im galaxies. The two distributions are similar.
The median values of R$_D^V$ and other disk parameters for our sample are given
in Table \ref{tab-aveexp}.
Similar values for the J-band fit
are also given in Table \ref{tab-aveexp}, and a comparison
of V-band and J-band R$_D$ is shown in Figure \ref{fig-rdjrdv}.
For most galaxies
the disk scale-length measured at J is close to that measured
at V (see also Bergvall \et\ 1999; Doublier \et\ 2001;
Cair\'os \et\ 2003; Noeske \et\ 2003), implying that the older stars and
younger stars trace the same disk structure.

Papaderos \et\ (1996) observed that the underlying disks in BCDs
have smaller R$_D$ by a factor of two compared to Im galaxies
of the same luminosity.
Patterson \& Thuan (1996) found that the dIms divide into two
groups: one group has twice the scale-length
of BCDs at the same luminosity and the other group has the same
scale-length.
A plot of R$_D^V$ as a function
of M$_{V_0}$ for our sample is shown in Figure \ref{fig-mvrdv}.
We see a separation of the Im and BCD galaxies, with the BCDs
having shorter scale-lengths. The Sm galaxies
fall between the Im galaxies and BCDs in scale-length at
a given galactic absolute magnitude.
We also
see a strong correlation between M$_V$ and R$_D^V$, first
pointed out by Hodge (1971) and seen in spirals (for example,
MacArthur \et\ 2004), in which lower luminosity
galaxies have smaller scale-lengths. We also see a separation
of systems in central surface brightness $\mu_0^V$, 
shown in the right panel of Figure \ref{fig-mvrdv}, as expected:
BCDs are small and high surface brightness, Ims are
lower in surface brightness and come in a range of sizes, and
Sms tend to be larger systems that are in between BCDs and
Ims in surface brightness.

\subsubsection{Comparison with spirals}

The exponential disks of a large sample of spirals have been
examined by de Jong (1996b), and it is useful to see how our sample
of galaxies compares to that sample.
In de Jong's Figure 3, reproduced here for V-band
as the top panel of Figure \ref{fig-galtype}, we see
that the average $\mu_0^V$ of the spiral galaxies
increases slightly with later galactic Hubble type,
with a scatter typically of 1--2 mag in 1 arcsec$^2$
(see also Roberts \& Haynes 1994).
The median values from the Sm, Im, and BCD samples is shown as the
horizontal line for the last 3 galaxy types in Figure \ref{fig-galtype}.
The median central surface brightnesses of our Sm and Im samples
continue the general trend seen in the spiral galaxies by de Jong
although the range of $\mu_0^V$ for the Im galaxies is over 6 magnitudes.
The BCD central surface brightnesses,
however, are considerably larger, 2.1 mag brighter on average
than the Im sample, and these
systems stand out as unusually bright.

De Jong's (1996) Figure 4, reproduced here for V-band
as the bottom panel of Figure \ref{fig-galtype},
shows the disk scale-lengths R$_D^V$
(see also Freeman 1970, Roberts \& Haynes 1994).
Again, the scatter is large for a given galaxy type,
and there is no obvious trend
with galactic Hubble type for the spirals.
The Im, BCD, and Sm samples all have median scale-lengths
that are smaller than those of the spirals, and BCDs are smaller
than the Ims which in turn are smaller than the Sms. But again,
the range is quite large within each type.

De Jong's (1996) Figure 5 shows a plot of scale-length versus
central surface brightness, reproduced here for V-band
as the right panel of Figure \ref{fig-mvrdv}.
Our systems lie to the lower left in this figure
with smaller scale-length and fainter central surface brightness.
As de Jong
points out, there is not much of a trend, and our irregular
galaxies just increase the scatter.
The left panel of Figure \ref{fig-mvrdv}, discussed in the previous
section, shows R$_D^V$ plotted as a function of galactic
M$_V$ with de Jong's spirals. The spirals and Im/BCD/Sm galaxies
produce a strong correlation (see also Freeman 1970).
Such a correlation is also seen
for a very large sample of luminous disk galaxies observed with the
Sloan Digital Sky Survey (SDSS; Blanton \et\ 2003).

Figure \ref{fig-mvmu0_primary} shows the distribution of
extrapolated central surface brightness $\mu_0^V$ versus galaxy magnitude
M$_{V_0}$ for all of the primary exponential disks in our survey. The dashed
line is the average Freeman (1970) value for spiral galaxies
(see also Phillipps \et\ 1987; van der Kruit 1987, 1989).
Freeman's $\mu_0^B=21.65$ mag arcsec$^{-2}$ has been corrected for reddening in the
same fashion that the spirals on this plot have been corrected
(E(B$-$V)$_i=0.3$)
and a (B$-$V)$_0=0.6$ is used to convert $\mu_0^B$ to $\mu_0^V$.
There is a lot of scatter, but overall the fainter galaxies have
fainter central surface brightnesses than the brighter galaxies
(see also de Jong \& van der Kruit 1994,
Roberts \& Haynes 1994, Burstein \et\ 1997).
On average they approach the Freeman value at around $M_{V_0}\sim-19$.
The BCDs have the highest $\mu_0^V$ and the Im types have the
lowest in our sample, as mentioned previously (see Fig.
\ref{fig-histmu0v}).
A correlation between luminosity and central surface brightness
is also seen for luminous disk galaxies observed with SDSS
(Blanton \et\ 2003).
From the slope of the correlation they conclude that the median
size of galaxies increases with luminosity, especially for the
most luminous galaxies.

\subsubsection{Complex surface brightness profiles}
\label{sect:complex}

While most of the sample galaxies have a V-band surface
brightness profile that is well fitted with a single exponential
disk, there are others that have more complex profiles.
The four types of variations that we see are
shown in Figures \ref{fig-steepouter_im} to \ref{fig-shallowouter_bcd}.
Each galaxy is offset
vertically for clarity; the major ticmarks on the ordinate
correspond to 2 mag arcsec$^{-2}$.
The bottom panels in each figure show B$-$V color profiles, with large
ticmarks corresponding to 0.2 mag and red increasing toward the top.

Of the Im galaxies, 23 (24\%)
have one of these complex types of profiles along with 10 (42\%) BCDs
and 8 (44\%) Sms.
The most common two-part profile (20 galaxies) is one in which
the light in the outer part of the galaxy drops more steeply
(Figures \ref{fig-steepouter_im} and \ref{fig-steepouter_sm}).
This was seen in two Im galaxies by Hidalgo \et\ (2003b), one
of which is also in our sample.
This type of behavior has also been seen in the outer parts
of spirals (van der Kruit \& Shostak 1982; Shostak \& Van der Kruit
1984; de Grijs \et\ 2001; Kregel \et\ 2002; Pohlen \et\ 2002;
Kregel \& van der Kruit 2004), including one
low luminosity spiral (Simon \et\ 2003).
(See also some of the surface brightness profiles of spirals shown by
Bell \et\ 2000 and MacArthur \et\ 2003).
It is now also seen in disks at high redshifts ($0.6<z<1.0$; P\'erez 2004).

Ten galaxies show a profile that is approximately flat over a
significant part of the galaxy, and
these galaxies are plotted in Figures \ref{fig-flatinner}
and \ref{fig-flatouter}.
This was also seen
for M81dwA and UGC 3817 by Patterson \& Thuan (1996).
DDO 68 is the only case in our sample where the
profile is nearly flat in the outer regions, but this feature was
also seen in the BCDs Tol 1214$-$277 and Tol 65 by Noeske \et\
(2003) and Cam 1148$-$2020 by Telles, Melnick, \& Terlevich
(1997).

Figures \ref{fig-shallowouter_im} and \ref{fig-shallowouter_bcd}
show profiles in which the light in the outer part of the
galaxy drops off less steeply compared to the inner regions. This is
seen in 11 galaxies: 8 BCDs, two Im galaxies with BCD-like
characteristics (IC 10 and NGC 3738),
and one normal Im (DDO 40).
This type of behavior is expected in BCDs and other
galaxies where the centrally concentrated star formation or
starburst steepens the surface brightness profile
there (Papaderos \et\ 1996; Cair\'os \et\ 2001; Noeske \et\ 2003).
The corresponding decrease in B$-$V for young stars in the central
steep part is seen in Figure \ref{fig-shallowouter_bcd}.  Unlike Doublier
\et\ (1999), we do not find BCDs that are better fit with an
r$^{1/4}$ law, although the average curvature in such a law, on a
surface-brightness versus radius plot, is the same as on our
double-exponential profiles where the outer part is relatively flat.

The division between the sub-sample with ``steep outer profiles''
and that with ``flat inner profiles'' is somewhat arbitrary.
There are those galaxies, such as DDO 75, DDO 168, DDO 185,
M81dwA, and UGC 8011,
with no surface brightness
gradient in the central regions. Then there are those,
such as DDO 69, Mrk 178, and UGCA 290, that have a short but noticeable
central gradient.
Finally, F651-2 has a
shallow, but more prolonged, gradient in the center.
It is arguable whether this last galaxy, in particular,
belongs in the ``flat inner profile'' group or the ``steep
outer profile'' group.
However, in our modelling of these profiles (Elmegreen \& Hunter 2005),
we combine these two groups together,
and so the exact line between them is not so important.

Figures \ref{fig-steepouter_im} to \ref{fig-flatinner}
show most of the double exponential
profiles that get steeper with radius.
The profiles are all
qualitatively the same, although those in Figure
\ref{fig-flatinner} are flatter in the inner regions.
The Sm galaxy colors tend to get bluer with radius at first
and then level off, while the Im galaxy colors are a little more
constant near the center and then get redder with distance. These
B$-$V color variations most likely reflect variations in the mean
stellar population ages and, possibly,
metallicities, with a small dependence on extinction
(\S\ref{sect:gradients}).

Statistics on where the break in the V-band profile R$_{Br}$ occurs
relative to R$_H$, R$_{1/2}^V$, and R$_D^V$ are given in
Table \ref{tab-twomu}. The R$_D^V$ used in this ratio
is the scale-length of the main disk: the inner
exponential for the profile that steepens in the outer parts
and the outer exponential for the
profile that is shallower in the outer parts.
We see that the average location of the profile break
relative to the Holmberg radius is similar whether the
profile steepens or becomes shallower in the outer parts.
However, a flat profile covers a somewhat smaller fraction
of the galaxy.

Relative to the disk scale-length, the break in the profile of Im
galaxies occurs on average at a radius of 1.5--1.7 scale-lengths
while that in the BCDs occurs on average at 1.9--2.2 scale-lengths.
This is shown in Figure \ref{fig-breakradii5}. Galaxies with
brighter central surface intensities tend to have larger ratios
$R_{Br}/R_D^V$, as noted by Pohlen et al. (2004)
and Kregel \& van der Kruit (2004) for spirals.
The break in the profiles of Pohlen
\et's (2002) three spirals occurs further out at 3.9$\pm$0.7
scale-lengths. Van der Kruit \& Shostak (1982) and Shostak \& van
der Kruit (1984) also found longer break radii---4.2 and 3.1
scale-lengths---for two spirals, and Kregel \& van der Kruit (2004)
found break radii (which they called truncation radii) at
$\sim$4 scale-lengths for a large group of high surface brightness
spirals.
However, de Grijs \et's (2001)
four spirals have breaks ranging from 1.6 to 3.3 scale-lengths,
similar to the irregulars observed here.
The disks measured at high redshift show breaks at about
1.8 scale-lengths (P\'erez 2004), more like the Im galaxies
than local spirals.

In Table \ref{tab-twomu} we also present the ratio of the
inner scale-length R$_{D,i}$ to outer
scale-length R$_{D,o}$ in V-band for the two-part exponentials.
In Pohlen \et's spirals this ratio is 2.0$\pm$0.2
which is
similar to the ratio for our Im and Sm galaxies with the steepening
exponential disk. Van der Kruit and Shostak's ratios bracket
Pohlen's ratio with values of 1.6 and 2.6.

What causes this break in the surface brightness profile? We
compared the locations of the break to the \ha\ surface brightness
profiles in order to see what was happening with the star
formation activity at that radius. We see that, of the 40 galaxies
with complex profiles, in 15 most of the detectable \ha\ ends
within one annulus of the location of the break. In another 18
galaxies, the \ha\ surface photometry drops more rapidly beginning
approximately around R$_{Br}$. Thus, in 83\% of these galaxies
there is some change in \sigha, and this would be a satisfying
explanation to the change in stellar surface brightness. However,
in 7 of the galaxies the \ha\ profile does not change or \ha\ ends
well before the break. Furthermore, there are 30 Ims (32\%), 11
(46\%) BCDs, and 1 (6\%) Sm that also exhibit a change in the drop
off rate of \sigha\, or \ha\ ends well before $\mu_{V_0}$, and yet they
do not show a break in their V-band surface brightness profiles.
Pohlen \et\ (2002) also could not find a convincing correlation
with \ha\ in spirals.

\ha\ is a measure of the current star formation, whereas
the broad-band colors integrate the star formation over longer
periods of time. Therefore, one might expect a better correlation
between the breaks in the surface brightness profiles and changes
in color profiles.
Among those galaxies with nearly flat inner profiles 5 of 10
show a change in B$-$V where the break in $\mu_{V_0}$ occurs.
All of these become redder beyond the break.
Of those with an exponential that becomes steeper,
7 become redder after the break, 1 becomes bluer, and 11 do not change.
The galaxy with the flat outer profile shows no change in color with
radius.

We have also compared the broad-band and \ha\ surface photometry
to the surface density of the neutral gas $\Sigma_{HI}$.
We plot $\mu_{V_0}$, \sigha, and $\Sigma_{HI}$ in
Figure \ref{fig-withhi} for all survey galaxies with
\HI\ surface density profiles $\Sigma_{HI}$
available from the literature (see the references listed in
the figure caption).
This includes three galaxies with breaks in their exponential profiles
(DDO 75, DDO 105, and DDO 168) and 9 that have single exponential profiles.
One can see that azimuthally-averaged
$\Sigma_{HI}$ has very little correspondence
with what is going on in the optical, including the breaks in
the exponential profiles, as well as changes in UBVJHK colors
(not shown in the figure but discussed in \S4.3.3).
Generally, $\Sigma_{HI}$ changes very slowly over the optical galaxy,
and when the change is more rapid, there is no correspondence with
optical changes.

Similarly, there is no correlation between where the break
in the surface brightness profile occurs and changes in
the \HI\ rotation curve of the galaxy (from the references in
the caption to Figure \ref{fig-withhi} and Swaters 1999).
The break occurs
after the rotation curve turns over in 5 galaxies,
during the solid body part of the rotation curve in 4 galaxies,
and about at the turn over in the rotation curve in 1 galaxy.
Five galaxies show no measurable rotation.

In a separate paper (Elmegreen \& Hunter 2005) we present a model
of star formation in outer galaxy disks that reproduces the
observed double exponential profile and the correlation between
break radius and galaxy magnitude. The model is based on the
concept that both gravitational instabilities and compression
trigger star formation, and that in the outer disk, beyond the
Kennicutt (1989) surface density threshold, only the compression
from low level turbulence and sporadic supernovae remain. Because
of this, the star formation rate in the outer disk does not drop
suddenly at the threshold, but has a more gradual,
exponential-like decline.

\section{On the Dalcanton \et\ (2004) study of Dust Lane Thickness}

Dalcanton \et\ (2004) proposed that large spiral galaxies have thin
dust and gas layers and small irregulars have thick dust and gas
layers or no obvious dust layers because the large galaxies
are more gravitationally unstable and this enhanced
instability makes the gaseous velocity
dispersions in those galaxies smaller.
While it may be that spirals are more
unstable than irregulars (\S1),
this is probably not the explanation for the difference
in gas layer thickness. As recognized by Dalcanton et al.,
thickness depends only on the velocity dispersion and
gravitational acceleration perpendicular to the plane, while stability
depends on the epicyclic frequency as well. Thus thin gas layers can be
in either stable or unstable disks, as can thick gas layers, depending on
the ratio of the perpendicular crossing time (which is the thickness
divided by the velocity dispersion) to the epicyclic time.  This ratio
is the Toomre stability parameter $Q$.

Figure \ref{fig-histsb25} suggests that the reason for the Dalcanton et al.\ (2004)
correlation between gas layer thickness and galaxy type is that
normal spirals have $\sim10$ times higher surface densities than
irregular galaxies, corresponding to a difference in surface
brightness of 2.8 magnitudes in 1 arcsec$^{-2}$.
In contrast, the velocity dispersions of the gas ($c$) do not
vary much among these different types, regardless of stability.
Thus, the gaseous scale height, which is
proportional to $c/\left(\pi G\Sigma\right)$, is high in irregulars because
the surface density is low. This is related to our previous
observation that the general interstellar medium pressure is lower in irregulars
than in spirals (Elmegreen \& Hunter 2000): the larger scale height makes the
average space density lower and the pressure lower for a given gas column
density and velocity dispersion.

In a disk composed of both gas and stars with small ratios of the
gaseous to stellar ($c_s$) velocity dispersions and masses, the gravitational
acceleration perpendicular to the plane is determined mostly
by stars and it increases linearly with height $z$  approximately as
$g(z)=zg_0/H_s$ for constant $g_0$ (this is an approximation to the
$\tanh(z/H_s)$ distribution of an isothermal stellar disk of scale height
$H_s$). The scale height $H$ for the gas is then given by $c^2/H\sim g(H)$, which
reduces to the relation $H=c\left(H_s/g_0\right)^{1/2}$.  It follows that
an approximate relation between the stellar and gaseous scale
heights and velocity dispersions is $H/H_s\sim c/c_s$.  If the gaseous
dispersion is somewhat uniform between galaxy types, and the
stellar dispersion is higher in large spirals (Dalcanton \et\ 2004),
then the ratio of the gaseous to the stellar
scale heights is smaller in spirals than in
irregulars, making the dust lanes look much thinner in spirals
compared to the background stellar disk.  The absolute gaseous scale height
is also larger in irregulars because the stellar disk thickness is larger
(\S\ref{sect:axes}).  All of these differences are unrelated
to the degree of disk stability.

Another reason dwarf galaxies have imperceptible dust lanes is that
the metallicity is low, making the extinction low per unit gas column
density, and the line of sight depth is low because of the small galaxy
size. Thus, the total opacity to dust is much lower in edge-on dwarfs than
in edge-on spirals.  An exception may be for the starburst cores of
BCDs, which could have a significant opacity because of the high
average gas density.

These points suggests that the Dalcanton et al.\ (2004) correlation
between dust lane thickness and galaxy rotation speed may be
explained in general terms by variations in the stellar surface
density and line-of-sight opacity. There is no obvious reason for
the reported suddenness of the transition at a rotation speed of
120 km s$^{-1}$, but it could
be from the combination of these two effects.  That is, the interstellar
medium
for an edge-on galaxy could become nearly optically thin between the
center and the edge at about the same rotation speed as the
average stellar surface density starts to drop by a factor of 10.
For example, this drop occurs in Figure \ref{fig-mvmu0_primary} at about
M$_{V_0}\sim=-19$ mag, and this corresponds to a rotation curve speed
of $\sim120$ km s$^{-1}$ from Figure 1 in Hunter, Hunsberger, \& Roye (2000).

\section{Conclusions} \label{sec-disc}

Observations of 136 Im, BCD, and Sm galaxies have been compiled to
provide a large sample for statistical studies of galaxy colors,
disk structures, and star formation processes in the most abundant
types of galaxies in the Universe.  The main results are:

1. The average and central surface brightnesses of Im
galaxies are generally lower than in spiral galaxies by a factor
of $\sim6$ ($\sim2$ mag in 1 arcsec$^{2}$). The extrapolated central
surface brightness for the main exponential part of the disk
begins to drop below the standard Freeman value for spirals at
$M_{V_0}$ fainter than $\sim-19$ mag. The low surface brightness
of Im galaxies makes their disks and dust lanes thicker than in
spirals, as observed by Dalcanton \et\ (2004).

2. Overall, Im galaxies are bluer than spirals.
However, they show a wide range of colors, complex color
gradients, and complex color patterns, indicating an irregular
mixture of stellar population ages, extinctions, and possibly metallicity
gradients.
Large-scale variations in colors are
clear indications of the manner in which star formation
can bubble around the galaxy on scales of kiloparsecs.

3. A normal fraction of Im galaxies are barred (23\% of our
Ims, 12\% of our BCDs, and 50\% or our Sms), but these bars
tend to be larger than in spirals, relative to the disk scale
length, sometimes occupying the entire bright part of the disk.
The bars are also clearly associated with an excess of star
formation in the Im galaxies.
The bars are not often offset from the center of the galaxy.

4. In spite of a bias against obviously interacting systems in our sample,
approximately one-third of the Im systems show large-scale morphological
peculiarities that may be an indication of past interactions.

5. The radial profiles of surface brightness are usually well
represented by an exponential over a large fraction of the disk,
but many galaxies show two exponentials or other complex patterns.
Some have flat profiles in the inner regions and a single
exponential beyond that, others have two exponentials with either
the inner one shallower or the inner one steeper. The cases with
steeper inner exponentials also have significant blue excesses
there, indicative of recently enhanced star formation in the
centers. The cases with shallow inner exponentials show no obvious
indications of their origin in peculiar star formation
patterns, color gradients, \HI\ profiles, or rotation curve features.
Star formation models that combine gravitational instabilities with
turbulence compression reproduce these profile features
(Elmegreen \& Hunter 2005).

\acknowledgments

We are deeply grateful to Ralph Nye who kept the Lowell
Observatory Hall 1.1 m and Perkins 1.8 m Telescopes and their
instruments operating and in excellent condition. We are also very
thankful to Phil Massey for obtaining some of the UBV images for
us and for doing the basic reductions of the Mosaic Imager images
that he obtained. We wish to thank Emily Bowsher for geometrically
matching some of the galaxies as part of the 2003 Research
Experiences for Undergraduates program of Northern Arizona
University funded by the National Science Foundation through grant
9988007. We also thank R. Shuping for obtaining several images as
part of KPNO Service Observing.
Funding for this work was provided by the Lowell Research
Fund and by the National Science Foundation through grants
AST-0204922 to DAH and AST-0205097 to BGE. This research has made
use of the NASA/IPAC Extragalactic Database (NED) which is
operated by the Jet Propulsion Laboratory, California Institute of
Technology, under contract with the National Aeronautics and Space
Administration.

Facilities: \facility{Lowell Observatory} \facility{KPNO}
\facility{CTIO}

\clearpage

\begin{deluxetable}{lllrrrcr}
\tabletypesize{\scriptsize}
\tablenum{1}
\tablecolumns{8}
\tablewidth{0pt}
\tablecaption{Galaxy Sample \label{tab-sample}}
\tablehead{
\colhead{} & \colhead{} & \colhead{}
& \colhead{D} & \colhead{} & \colhead{}
& \colhead{$\log$ M$_{gas}$\tablenotemark{e}} & \colhead{} \\
\colhead{Galaxy} & \colhead{Other Names\tablenotemark{a}}
& \colhead{Type\tablenotemark{b}}
& \colhead{(Mpc)} & \colhead{Ref.\tablenotemark{c}}
& \colhead{E(B$-$V)$_f$\tablenotemark{d}}
& \colhead{(M\solar)} & \colhead{Ref.\tablenotemark{f}}
}
\startdata
\cutinhead{Im Galaxies}
A1004$+$10\dotfill & PGC 29428, UGC 5456, IRAS F10046$+$1036  & I0? &  6.5\x & \nodata & 0.01\xx & 7.59 & 23 \\
A2228$+$33\dotfill & PGC 69019, UGC 12060, IRAS F22282$+$3334 & IBm & 16.9\x & \nodata & 0.01\xx & 9.35 & 22 \\
CVnIdwA\dotfill & UGCA 292                                    & Im? &  4.1\x & 27      & 0.01\xx & 7.92 & 51 \\
D508-2\dotfill  & LSBC F508-V01                               & Im  & 29.9\x & \nodata & 0.00\xx & 9.18 & 42 \\
D575-5\dotfill  & LSBC F575-03                                & dI  &  6.2\x & \nodata & 0.03\xx & 7.52 & 42 \\
\enddata
\tablecomments{Table 1 is published
in its entirety in the electronic edition
of the {\it Astrophysical Journal}. A portion is shown here
for guidance regarding its form and content.}
\tablenotetext{a}{Selected alternate identifications
obtained from NED.}
\tablenotetext{b}{Morphological Hubble types are from de Vaucouleurs et al.\ (1991). If no type is
given there, we have used types given by NED.}
\tablenotetext{c}{Reference for the distance to the galaxy.
If no reference is given, the distance
was determined from V$_{GSR}$ given by de Vaucouleurs et al.\ (1991) and a Hubble constant
of 65 km s$^{-1}$ Mpc$^{-1}$.}
\tablenotetext{d}{Foreground reddening from Burstein \& Heiles (1984).}
\tablenotetext{e}{The gas mass is the \protect\HI\ mass M$_{HI}$ plus 1.34$\times$M$_{HI}$
 to account for He.}
\tablenotetext{f}{Reference from which the \protect\HI\ mass was taken.
Masses were modified to reflect the distances used here as necessary.}
\tablerefs{
(1) Allsopp 1978;
(2) Aparicio 1994;
(3) Aparicio, Tikhonov, \& Karachentsev 2000;
(4) Bottinelli \et\ 1990;
(5) Bureau \& Carignan 2002;
(6) Carignan \& Beaulieu 1989;
(7) de Blok, McGaugh, \& van der Hulst 1996;
(8) de Blok \& Walter 2000;
(9) Dohm-Palmer \et\ 1998;
(10) Dolphin 2000;
(11) Dolphin \et\ 2002;
(12) Dolphin \et\ 2003;
(13) Fisher \& Tully 1981;
(14) Freedman \et\ 2001;
(15) Gallagher \et\ 1998;
(16) Gallart, Aparicio, \& Vilchez 1996c;
(17) Gordon \& Gottesman 1981;
(18) Hidalgo, Mar\'in-Franch, \& Aparicio 2003a;
(19) Huchtmeier, Hopp, \& Kuhn 1997;
(20) Huchtmeier, Seiradakis, \& Materne 1981;
(21) Hunter 2001;
(22) Hunter \& Gallagher 1985b;
(23) Hunter, Gallagher, \& Rautenkranz 1982;
(24) Karachentsev, Aparicio, \& Makarova  1999;
(25) Karachentsev, Musella, \& Grimaldi 1996;
(26) Karachentsev \et\ 2002;
(27) Karachentsev \et\ 2003a;
(28) Karachentsev \et\ 2003b;
(29) Kniazen \et\ 2000;
(30) Lee \et\ 1999;
(31) Lee \& Kim 2000;
(32) Lo, Sargent, \& Young 1993;
(33) Maiz-Apellaniz, Cieza, \& Mackenty 2002;
(34) Makarova \et\ 1998;
(35) Massey \& Armandroff 1995;
(36) M\'endez \et\ 2002;
(37) Meurer, Staveley-Smith, \& Killeen 1998;
(38) Miller \et\ 2001;
(39) Minniti \& Zijlstra 1997;
(40) Nordgren \et, 2003;
(41) O'Connell \et\ 1994;
(42) Pildis, Schombert, \& Eder 1997;
(43) RC3;
(44) Sargent, Sancisi, \& Lo 1983;
(45) Stil \& Israel 2002;
(46) Swaters 1999;
(47) Thuan \& Martin 1981;
(48) Tolstoy \et\ 1995;
(49) Tosi \et\ 2001;
(50) van Zee, Haynes, \& Giovanelli 1995;
(51) van Zee \et\ 1997b;
and
(52) Young \& Lo 1997.
}
\end{deluxetable}

\clearpage

\begin{deluxetable}{llcccccrrcrrc}
\tabletypesize{\scriptsize}
\rotate
\tablenum{2}
\tablecolumns{13}
\tablewidth{0pt}
\tablecaption{Observations and Photometry Parameters \label{tab-obs}}
\tablehead{
\colhead{} & \colhead{} & \colhead{} & \colhead{} & \colhead{} & \colhead{}
& \colhead{}
& \multicolumn{5}{c}{Ellipse photometry parameters\tablenotemark{a}}
& \colhead{} \\
\cline{8-12}
\colhead{} & \colhead{} & \colhead{} &\colhead{}
& \colhead{Exposures\tablenotemark{c}}
& \colhead{Seeing\tablenotemark{d}}
& \colhead{Scale}
& \colhead{P.A.}
& \colhead{} & \colhead{Step} & \multicolumn{2}{c}{Center (J2000)}
& \colhead{Calib rms\tablenotemark{e}} \\
\colhead{Galaxy}
& \colhead{Date}
& \colhead{Instr.\tablenotemark{b}}
& \colhead{Filters}
& \colhead{(s)}
& \colhead{(arcsec)}
& \colhead{(arcsec)}
& \colhead{(deg)}
& \colhead{{\it b/a}} & \colhead{(arcsec)}
& \colhead{R.A.} & \colhead{Decl.}
& \colhead{(mag)}
}
\startdata
\cutinhead{Im Galaxies}
A1004+10\dotfill  & 9903 & LO1.8m &  BV & $3\times1200$,$3\times900$ & 1.7,1.5 & 0.61 & $-$33.3 & 0.63 &  9.1 & 10 07 19.6 &  10 21 47 & 0.07,0.04 \\
A2228+33\dotfill  & 9810 & LO1.1m & BV & $3\times2400$,$3\times1200$ & 3.6,3.0 & 1.13 & 38.5 & 0.92 & 11.3 & 22 30 34.0 &  33 49 14 & 0.03,0.03 \\
                  & 9910 & LO1.1m & U   & $3\times2400$ & 3.4 & 1.13 & \nodata & \nodata & \nodata  & \nodata & \nodata & 0.05 \\
CVnIdwA\dotfill   & 0005 & LO1.1m & UBV & $3\times1800$,$8\times900$,$9\times600$ & 3.8,3.9,2.8 & 1.13 & 79.5 & 0.78 & 11.3 & 12 38 40.2 &  32 45 40 & 0.04,0.03,0.02 \\
D508-2\dotfill    & 0004 & LO1.1m & UBV & $4\times1800$,$3\times1800$,$7\times1200$ & 3.8,3.8,3.1 & 1.13 & 70.6 & 0.76 &  9.1 & 13 04 34.3 &  26 46 24 & 0.07,0.02,0.02 \\
D575-5\dotfill    & 9903 & LO1.8m &  BV & $4\times1800$,$4\times900$ & 3.3,3.1 & 0.61 & 56.0 & 0.51 & 9.1 & 12 55 41.0 & 19 12 35 & 0.03,0.04 \\
                  & 9904 & LO1.8m & U   & $2\times1800$ & 3.1 & 0.61 & \nodata & \nodata & \nodata & \nodata & \nodata & 0.05 \\
\enddata
\tablecomments{Table 2 is published
in its entirety in the electronic edition
of the {\it Astrophysical Journal}. A portion is shown here
for guidance regarding its form and content.}
\tablenotetext{a}{Position angle P.A., minor-to-major axis ratio
$b/a$, ellipse semi-major axis step size,
and position of center used to do photometry in concentric
ellipses. The P.A. given here has been corrected for the P.A. of the CCD camera
on the sky, usually about 0.5\arcdeg, but the P.A.
appropriate to the image itself was used for the photometry.}
\tablenotetext{b}{Telescope used for the observations:
LO1.8m$=$1.8 m Perkins Telescope at Lowell Observatory;
LO1.1m$=$1.1 m Hall Telescope at Lowell Observatory;
KPNO4m$=$Kitt Peak National Observatory 4 m Telescope;
CTIO1.5m$=$Cerro Tololo Inter-American Observatory 1.5 m Telescope;
CTIOSc$=$Cerro Tololo Inter-American Observatory Schmidt Telescope;
CTIO4m$=$Cerro Tololo Inter-American Observatory 4 m Telescope.
}
\tablenotetext{c}{A () designates images taken under clear sky conditions
to calibrate the other frames taken under non-photometric conditions.}
\tablenotetext{d}{FWHM of a stellar profile on the final combined
image.}
\tablenotetext{e}{Photometric calibration rms for the filters
listed in Column 4.}
\end{deluxetable}

\clearpage

\begin{deluxetable}{lrrrrrrrrrrrrr}
\tabletypesize{\scriptsize}
\rotate
\tablenum{3}
\tablecolumns{14}
\tablewidth{0pt}
\tablecaption{Integrated Photometry \label{tab-integrated}}
\tablehead{
\colhead{} & \colhead{R\tablenotemark{b}}
& \colhead{} & \colhead{} & \colhead{} & \colhead{}
& \colhead{} & \colhead{} & \colhead{} & \colhead{} & \colhead{}
& \colhead{} & \colhead{} & \colhead{} \\
\colhead{~~~~~Galaxy~~~~~}
& \colhead{(arcmin)}
& \colhead{M$_{V_0}$}
& \colhead{$\sigma_{M_{V}}$}
& \colhead{(U$-$B)$_0$}
& \colhead{$\sigma_{U-B}$}
& \colhead{(B$-$V)$_0$}
& \colhead{$\sigma_{B-V}$}
& \colhead{(V$-$J)$_0$}
& \colhead{$\sigma_{V-J}$}
& \colhead{(J$-$H)$_0$}
& \colhead{$\sigma_{J-H}$}
& \colhead{(H$-$K)$_0$}
& \colhead{$\sigma_{H-K}$}
}
\startdata
\cutinhead{Im Galaxies}
A1004+10\dotfill  &    0.33 & $-$15.178 & 0.001 & \nodata & \nodata &  0.286 & 0.001 & \nodata & \nodata & \nodata & \nodata & \nodata & \nodata \\
                    &  1.37 & $-$15.930 & 0.001 & \nodata & \nodata &  0.326 & 0.001 & \nodata & \nodata & \nodata & \nodata & \nodata & \nodata \\
A2228+33\dotfill  &    0.70 & $-$17.002 & 0.008 & $-$0.062 & 0.016 &  0.502 & 0.012 & \nodata & \nodata & \nodata & \nodata & \nodata & \nodata \\
                    &  2.46 & $-$17.754 & 0.013 & $-$0.098 & 0.027 &  0.510 & 0.020 & \nodata & \nodata & \nodata & \nodata & \nodata & \nodata \\
CVnIdwA\dotfill   &    0.56 & $-$11.902 & 0.070 & $-$0.675 & 0.102 &  0.096 & 0.091 & \nodata & \nodata & \nodata & \nodata & \nodata & \nodata \\
                    &  1.32 & $-$12.655 & 0.083 & $-$0.551 & 0.142 &  0.207 & 0.111 & \nodata & \nodata & \nodata & \nodata & \nodata & \nodata \\
D508-2\dotfill    &    0.32 & $-$14.988 & 0.063 & $-$0.661 & 0.116 & $-$0.127 & 0.090 & \nodata & \nodata & \nodata & \nodata & \nodata & \nodata \\
                    &  0.76 & $-$15.742 & 0.072 & $-$0.650 & 0.130 & $-$0.127 & 0.101 & \nodata & \nodata & \nodata & \nodata & \nodata & \nodata \\
D575-5\dotfill    &    0.49 & $-$12.100 & 0.034 & $-$0.282 & 0.055 &  0.159 & 0.039 & \nodata & \nodata & \nodata & \nodata & \nodata & \nodata \\
                    &  0.91 & $-$12.852 & 0.032 & $-$0.137 & 0.053 &  0.199 & 0.037 & \nodata & \nodata & \nodata & \nodata & \nodata & \nodata \\
\enddata
\tablecomments{Table 3 is published
in its entirety in the electronic edition
of the {\it Astrophysical Journal}. A portion is shown here
for guidance regarding its form and content.}
\tablenotetext{a}{These values are V$-$H rather than V$-$J.}
\tablenotetext{b}{Integrated photometry is given at R$_{1/2}^V$ and the
total extent of the V-band image, as well as the total extents in
JHK.}
\end{deluxetable}

\clearpage

\voffset=0.5truein
\begin{deluxetable}{lrrrcrrrrrrrrrr}
\tabletypesize{\scriptsize}
\rotate
\tablenum{4}
\tablecolumns{15}
\tablewidth{0pt}
\tablecaption{Structural Parameters\label{tab-structure}}
\tablehead{
\colhead{}
& \multicolumn{4}{c}{Primary\tablenotemark{a}}
& \colhead{}
& \colhead{}
& \multicolumn{4}{c}{Secondary\tablenotemark{a}}
& \colhead{}
& \colhead{}
& \colhead{}
& \colhead{} \\
\cline{2-5} \cline{8-11}
\colhead{}
& \colhead{R$_D$}
& \colhead{$\sigma_{R_D}$}
& \colhead{$\mu_0$}
& \colhead{$\sigma_{\mu_0}$}
& \colhead{}
& \colhead{R$_{Br}$}
& \colhead{R$_D$}
& \colhead{$\sigma_{R_D}$}
& \colhead{$\mu_0$}
& \colhead{$\sigma_{\mu_0}$}
& \colhead{R$_{25}$\tablenotemark{c}}
& \colhead{R$_H$\tablenotemark{c}}
& \colhead{R$_{1/2}^V$\tablenotemark{c}}
& \colhead{$\sigma_R$\tablenotemark{d}}\\
\colhead{Galaxy}
& \colhead{(kpc)}
& \colhead{(kpc)}
& \colhead{(mag arcsec$^{-2}$)}
& \colhead{(mag arcsec$^{-2}$)}
& \colhead{Sec?\tablenotemark{b}}
& \colhead{(kpc)}
& \colhead{(kpc)}
& \colhead{(kpc)}
& \colhead{(mag arcsec$^{-2}$)}
& \colhead{(mag arcsec$^{-2}$)}
& \colhead{(kpc)}
& \colhead{(kpc)}
& \colhead{(kpc)}
& \colhead{(kpc)}
}
\startdata
\cutinhead{Im Galaxies}
           A1004+10\dotfill  &    0.34 &  0.01 & 20.37 &  0.12 & N & \nodata & \nodata & \nodata & \nodata & \nodata &  1.28 &  1.83 &  0.62& 0.08 \\
           A2228+33\dotfill  &    2.56 &  0.11 & 22.85 &  0.09 & N & \nodata & \nodata & \nodata & \nodata & \nodata &  3.82 &  7.39 &  3.44& 0.23 \\
           CVnIdwA\dotfill   &    0.64 &  0.15 & 24.30 &  0.26 & N & \nodata & \nodata & \nodata & \nodata & \nodata & \nodata & 1.04 & 0.67& 0.06 \\
           D508-2\dotfill    &    1.80 &  0.17 & 24.02 &  0.21 & N & \nodata & \nodata & \nodata & \nodata & \nodata &  1.93 &  4.80 &  2.80& 0.36 \\
           D575-5\dotfill    &    0.85 &  0.12 & 24.69 &  0.15 & N & \nodata & \nodata & \nodata & \nodata & \nodata & \nodata & 1.27 & 0.89& 0.07 \\
\enddata
\tablewidth{600pt}
\tablecomments{Table 4 is published
in its entirety in the electronic edition
of the {\it Astrophysical Journal}. A portion is shown here
for guidance regarding its form and content.}
\tablenotetext{a}{Entries on the first line for a galaxy are measured
from the
V-band. Entries on a second line, if present, are \\
measured from the J-band
image.
The exception are the ``Other Objects,'' which are measured only in J-band.}
\tablenotetext{b}{If the surface photometry was fit with two parts
it is noted
here as inward of the primary fit (``I'') or outward (``O'').  The \\
parameter
R$_{Br}$ is the radius at which the two fits cross each other.
An ``N'' means that the surface photometry was fit \\
with only one exponential.}
\tablenotetext{c}{R$_{25}$ and R$_H$ are measured from reddening-corrected
B-band surface photometry. R$_{1/2}^V$ is measured from the \\
reddening-corrected
V-band surface photometry.}
\tablenotetext{d}{The uncertainty in R$_{25}$, R$_H$, and R$_{1/2}^V$ is
one-quarter of the annulus width in the ellipse photometry from which these quantities \\
were determined. A minimum of 2.5\arcsec\ is imposed, comparable to the average seeing.}
\end{deluxetable}

\clearpage

\begin{deluxetable}{lrccccccccccccc}
\tabletypesize{\scriptsize}
\rotate
\tablenum{5}
\tablecolumns{14}
\tablewidth{0pt}
\tablecaption{Disk Characteristics\label{tab-stuff}}
\tablehead{
\colhead{}
& \colhead{}
& \colhead{}
& \colhead{}
& \multicolumn{5}{c}{Bar\tablenotemark{d}}
& \colhead{$\mu_{25}$\tablenotemark{f}}
& \colhead{$\sigma_{\mu_{25}}$}
& \colhead{$\mu_D^V$\tablenotemark{f}}
& \colhead{$\sigma_{\mu_D^V}$}
& \colhead{$\mu_{2.5D}$\tablenotemark{f}}
& \colhead{$\sigma_{\mu_{2.5D}}$} \\
\cline{5-9}
\colhead{}
& \colhead{$i$\tablenotemark{a}}
& \colhead{Complex}
& \colhead{}
& \colhead{R$_{Bar}$}
& \colhead{}
& \colhead{}
& \colhead{$\Delta$P.A.}
& \colhead{}
& \colhead{(mag}
& \colhead{(mag}
& \colhead{(mag}
& \colhead{(mag}
& \colhead{(mag}
& \colhead{(mag} \\
\colhead{Galaxy}
& \colhead{(deg)}
& \colhead{$\mu_V$?\tablenotemark{b}}
& \colhead{Pec?\tablenotemark{c}}
& \colhead{(kpc)}
& \colhead{R$_{Bar}$/R$_D^V$}
& \colhead{$b/a$}
& \colhead{(deg)}
& \colhead{$\Delta$R/R$_D^V$\tablenotemark{e}}
& \colhead{arcsec$^{-2}$)}
& \colhead{arcsec$^{-2}$)}
& \colhead{arcsec$^{-2}$)}
& \colhead{arcsec$^{-2}$)}
& \colhead{arcsec$^{-2}$)}
& \colhead{arcsec$^{-2}$)}
}
\startdata
\cutinhead{Im Galaxies}
A1004+10\dotfill  &   54 & \nodata & C       & \nodata & \nodata & \nodata & \nodata & \nodata &  22.86  & 0.00    &  21.13  & 0.00 &   21.93 &    0.00 \\
A2228+33\dotfill  &   24 & \nodata & \nodata &  2.18$\pm$0.43   & 0.8     & 0.69    & 78      & 0.00       &  24.12  & 0.01    &  23.17  & 0.00 &   24.29 &    0.01 \\
CVnIdwA\dotfill   &   40 & \nodata & \nodata & \nodata & \nodata & \nodata & \nodata & \nodata & \nodata & \nodata &  25.05  & 0.07 & \nodata & \nodata \\
D508-2\dotfill    &   42 & \nodata & \nodata & \nodata & \nodata & \nodata & \nodata & \nodata &  24.80  & 0.08    &  24.89  & 0.08 &   25.55 &    0.06 \\
D575-5\dotfill    &   64 & \nodata & \nodata & \nodata & \nodata & \nodata & \nodata & \nodata & \nodata & \nodata &  25.43  & 0.03 & \nodata & \nodata \\
\enddata
\tablewidth{600pt}
\tablecomments{Table 5 is published
in its entirety in the electronic edition
of the
{\it Astrophysical Journal}. A portion is shown here \\
for guidance regarding its form and content.}
\tablenotetext{a}{The inclination of the galaxy, determined from the $b/a$ in
Table 2 under the assumption
that $(b/a)_0=0.3$ for the irregulars
and 0.2 for the spirals.}
\tablenotetext{b}{If a galaxy has a complex $\mu_V$, it is marked as
FI$=$``flat inner part'', FO$=$``flat outer part'',
SO$=$``steeper outer part'', SI$=$``steeper inner part.''}
\tablenotetext{c}{``M'' indicates the presence of a morphological peculiarity,
and ``C'', the presence of a color peculiarity.}
\tablenotetext{d}{Characteristics of a bar structure, if present, are given.
R$_{Bar}$ is the semi-major axis
length of the bar. This quantity is followed
by a ``?'' if there is some question as to
whether what is seen is a bar structure. The uncertainty is the maximum amount
by which we feel that the bar length could reasonably be altered based
on contour plots.
$b/a$ is the minor to major axis ratio of the bar.
$\Delta$P.A. is the difference in the position angle between the bar
and the outer galaxy. If no value is given, the outer galaxy was too
round to reliably determine the P.A.}
\tablenotetext{e}{Radial separation between center of bar and center of
outer isophotes in the plane of the galaxy as determined in the V-band
relative to the scale-length R$_D^V$ of the disk.
An offset less than the seeing of the boxcar-smoothed image was set to zero.}
\tablenotetext{f}{$\mu_{25}$, $\mu_D^V$, and $\mu_{2.5D}$ are the average surface brightnesses,
magnitudes in 1 arcsec$^2$,
within the radii R$_{25}$, R$_D^V$, and 2.5$\times$R$_D^V$, respectively. $\mu_{25}$ is measured on the B-band
image, and $\mu_D^V$ and $\mu_{2.5D}$, on the V-band image. The photometry is normalized to
the circular area, that is, $\pi$R$^2$.}
\end{deluxetable}
\clearpage

\voffset=0.5truein
\begin{deluxetable}{lccccrccccrc}
\tabletypesize{\scriptsize}
\tablenum{6}
\tablecolumns{12}
\tablewidth{0pt}
\tablecaption{Color gradients\label{tab-colgrad}}
\tablehead{
\colhead{}
& \colhead{}
& \colhead{}
& \multicolumn{4}{c}{Part 1\tablenotemark{a}}
& \colhead{}
& \multicolumn{4}{c}{Part 2\tablenotemark{a}} \\
\cline{4-7} \cline{9-12}
\colhead{}
& \colhead{}
& \colhead{}
& \colhead{R$_{beg}$\tablenotemark{b}}
& \colhead{R$_{end}$\tablenotemark{b}}
& \colhead{Gradient}
& \colhead{$\sigma_{grad}$}
& \colhead{}
& \colhead{R$_{beg}$\tablenotemark{b}}
& \colhead{R$_{end}$\tablenotemark{b}}
& \colhead{Gradient}
& \colhead{$\sigma_{grad}$} \\
\colhead{Galaxy}
& \colhead{Color}
& \colhead{}
& \colhead{(kpc)}
& \colhead{(kpc)}
& \colhead{(mag/kpc)}
& \colhead{(mag/kpc)}
& \colhead{}
& \colhead{(kpc)}
& \colhead{(kpc)}
& \colhead{(mag/kpc)}
& \colhead{(mag/kpc)}
}
\startdata
\cutinhead{Im Galaxies}
A1004+10 \dotfill & B$-$V & &  0.00 &  1.01 & 0.0      &  0.0   & & 1.01 &  2.44 &  0.339  &  0.026 \\
DDO 24   \dotfill & B$-$V & &  0.00 &  3.86 & $-$0.071 &  0.002 & & \nodata & \nodata & \nodata & \nodata \\
                  & U$-$B & &  0.00 &  3.27 & $-$0.060 &  0.012 & & \nodata & \nodata & \nodata & \nodata \\
DDO 25   \dotfill & B$-$V & &  0.00 &  4.54 & $-$0.034 &  0.005 & & \nodata & \nodata & \nodata & \nodata \\
DDO 34   \dotfill & U$-$B & &  0.00 &  2.65 & $-$0.200 &  0.016 & & \nodata & \nodata & \nodata & \nodata \\
DDO 38   \dotfill & B$-$V & &  0.00 &  5.61 & $-$0.052 &  0.004 & & \nodata & \nodata & \nodata & \nodata \\
                  & U$-$B & &  0.00 &  5.61 & $-$0.074 &  0.015 & & \nodata & \nodata & \nodata & \nodata \\
\enddata
\tablecomments{Table 6 is published
in its entirety in the electronic edition
of the {\it Astrophysical Journal}. A portion is shown here
for guidance regarding its form and content.}
\tablenotetext{a}{Surface photometry profiles for some galaxies were
fit with two gradients with different slopes.}
\tablenotetext{b}{The range of radii over which the gradient was measured.}
\end{deluxetable}

\clearpage
\voffset=0truein

\begin{deluxetable}{lccc}
\tablenum{7}
\tablecaption{Median disk parameters. \label{tab-aveexp}}
\tablewidth{0pt}
\tablehead{
\colhead{Parameter}
& \colhead{Im}
& \colhead{BCD}
& \colhead{Sm}
}
\startdata
$\mu_0^V$ (mag of 1 arcsec$^2$) & 23.1 & 21.0 & 22.2 \\
$\mu_0^J$ (mag of 1 arcsec$^2$) & 21.2 & 19.9 & 20.2 \\
R$_D^V$ (kpc)                   & 1.0 & 0.5 & 1.7 \\
R$_D^J$ (kpc)                   & 0.7 & 0.4 & 1.3 \\
$b/a$                           & 0.6 & 0.6 & 0.7 \\
M$_{V_0}$                       & $-15.4$ & $-16.0$ & $-17.4$ \\
R$_{1/2}^V$ (kpc)                 & 1.4 & 0.8 & 3.0 \\
$\mu_{25}$ (mag of 1 arcsec$^2$) & 24.4 & 23.1 & 24.1 \\
$\mu_D^V$ (mag of 1 arcsec$^2$)  & 24.0 & 21.4 & 23.0 \\
\enddata
\end{deluxetable}

\clearpage

\begin{deluxetable}{lccccccc}
\tablenum{8}
\tablecaption{Average characteristics of two-part surface brightness profiles.
\label{tab-twomu}}
\tablewidth{0pt}
\tablehead{
\colhead{Type}
& \colhead{No.}
& \colhead{Percent}
& \colhead{R$_{Br}$/R$_H$}
& \colhead{R$_{Br}$/R$_{1/2}^V$}
& \colhead{R$_{Br}$/R$_D^V$\tablenotemark{a}}
& \colhead{R$_{D,i}^V$/R$_{D,o}^V$\tablenotemark{b}}
}
\startdata
\cutinhead{Exponential becomes steeper}
Im & 11 & 12 & 0.8$\pm$0.3 & 1.4$\pm$0.3 & 1.5$\pm$0.8 & 2.5$\pm$0.6 \\
Sm &  8 & 44 & 0.7$\pm$0.1 & 1.5$\pm$0.1 & 2.2$\pm$0.5 & 1.9$\pm$0.3 \\
\cutinhead{Exponential becomes shallower}
Im &  3 &  3 & 0.6$\pm$0.3 & 1.3$\pm$0.5 & 1.7$\pm$1.0 & 0.6$\pm$0.2 \\
BCD & 8 & 33 & 0.6$\pm$0.1 & 1.9$\pm$0.3 & 2.0$\pm$0.5 & 0.6$\pm$0.1 \\
\cutinhead{Flat inner profile}
Im &  8 &  9 & 0.5$\pm$0.1 & 0.8$\pm$0.2 & 1.6$\pm$0.7 & \nodata \\
BCD & 2 &  8 & 0.4$\pm$0.3 & 0.8$\pm$0.4 & 1.9$\pm$1.3 & \nodata \\
\cutinhead{Flat outer profile}
Im &  1 &  1 & 0.4$\pm$0.3 & 0.8$\pm$0.4 & 1.9$\pm$1.3 & \nodata \\
\enddata
\tablenotetext{a}{R$_D^V$ is the scale-length of the inner V-band profile for the
exponential that becomes steeper in the outer parts
and of the outer profile
for the exponential that is shallower in the outer parts.}
\tablenotetext{b}{Ratio of inner scale-length R$_{D,i}^V$ to the outer
scale-length R$_{D,o}^V$.}
\end{deluxetable}

\clearpage
\begin{figure}
\epsscale{0.8}
\caption{False-color display of an example of the dataset
using the Im galaxy NGC 2366.
Upper: The logarithms of the V-band and H$\alpha$ images are shown 
in order to allow comparison of faint outer features and bright
inner features.
Lower left: A composite of the UBV images, with V as red, B as green,
and U as blue.
Lower right: A composite of the JHK images, with K as red, H as green,
and J as blue.
North is at the top; East is to the left.
\label{fig-images}}
\end{figure}

\begin{figure}
\caption{
Azimuthally-averaged V-band, J-band, and H$\alpha$ surface photometry
of NGC 2366, an example of our surface photometry data set.
All are corrected for reddening.
The scales for $\Sigma_{H\alpha}$, $\mu_{V_0}$, and $\mu_{J_0}$
have been set
so that they cover the same logarithmic interval.
The solid lines are
fits to the V and J-band surface photometry.
The radii corresponding to \protect\rtf\ and \protect\rh\ are marked
with vertical lines near the bottom of the plot.
The deviations in the surface photometry of NGC 2366 near 1 kpc radius are
due to the supergiant \protect\HII\ region NGC 2363.
\label{fig-sb}}
\end{figure}

\begin{figure}
\epsscale{0.8}
\caption{As for Figure
\protect\ref{fig-sb}, but showing the azimuthally-averaged colors.
The solid line in each panel is the average color of points with
relatively low uncertainties.
The deviations in the colors of NGC 2366 near 1 kpc radius are
due to the supergiant \protect\HII\ region NGC 2363.
\label{fig-colors}}
\end{figure}

\begin{figure}
\epsscale{0.55}
\caption{Cuts along the major axis through the V-band image of DDO 48
illustrate the sky background and the location of
the breaks in the surface photometry profile without
azimuthal averaging. In the top panel a cut through the sky-subtracted
image is shown; the cut sums over 11 pixels (12.5\arcsec) and then
is averaged along the cut in 10 pixel increments.
The magnitude is on an arbitrary scale,
formed from the counts in the image plus a constant of 25.
The vertical lines mark the center of the image (dashed line)
and the break radius (solid lines).
The solid sloped lines are the fits to the azimuthally-averaged surface photometry
used elsewhere in this paper, and
anchored on this plot at the first point in the radial interval of that
portion of the double exponential.
The radial interval plotted here also corresponds to that plotted in the
azimuthally-averaged surface brightness plot of this galaxy.
The bottom panel shows a similar cut through the original V-band image, and
the dashed line is this cut through the two-dimensional fit to the sky.
These are plotted as counts rather than as magnitudes, and
here the cut spans the entire V-band image.
The small bottom-most panel is the cut through the sky-subtracted
image, shown as counts rather than as magnitudes and displayed to
view the background level after sky-subtraction. The portion of the
galaxy corresponding to the outer exponential stands clearly above the
sky noise.
\label{fig-cuts}}
\end{figure}

\clearpage

\begin{figure}
\epsscale{0.8}
\caption{ Number distribution of the survey
galaxies in integrated M$_{V}$, corrected for reddening. The
vertical dashed line marks the median value in M$_{V_0}$ for the
Im galaxies. Spiral galaxies from the sample of Kennicutt (1983)
are shown for comparison.
\label{fig-histmv}}
\end{figure}

\begin{figure}
\epsscale{0.8}
\caption{ Number distribution of the survey
galaxies in $\mu_{25}$, the average surface brightness within a
B-band isophote of 25 mag in one arcsec$^2$, corrected for
reddening. The dashed vertical line marks the median value of
$\mu_{25}$ for the Im group. Spiral galaxies from the sample of
Kennicutt (1983) are shown for comparison and have been corrected
to normalization by the circular area $\pi R^2$.
\label{fig-histsb25}}
\end{figure}

\begin{figure}
\caption{ Number distribution of the survey
galaxies in $\mu_D^V$, the average surface brightness within
R$_D^V$ in V-band, corrected for reddening. The dashed vertical
line marks the median value of $\mu_D^V$ for the Im group.
\label{fig-histsbdv}}
\end{figure}

\begin{figure}
\caption{ Number distribution of the
survey galaxies in $R_{1/2}^V$, the radius that contains half of
the light of the galaxy in V-band. The dashed vertical line marks
the median value of $R_{1/2}^V$ for the Im group.
\label{fig-histhalfr}}
\end{figure}

\clearpage

\begin{figure}
\caption{ Number distribution of the
barred survey galaxies in semi-major axis length of the bar R$_{Bar}$
relative to the disk scale-length R$_D^V$.
\label{fig-rbrd}}
\end{figure}

\begin{figure}
\caption{Profile cuts along the major and minor axes of the bars
in 6 of the survey galaxies.
The three galaxies in the top row are the only barred systems in
our survey showing a flat bar profile. The three galaxies in
the bottom row are examples of the rest of the barred systems
which show an exponential bar profile.
The plots are centered on the center
of the bar as determined in the V-band image.
The cuts average over 5\protect\arcsec. The $\mu_V$
have not been corrected for reddening.
\label{fig-barcuts}}
\end{figure}

\begin{figure}
\caption{Comparison of properties for barred (hatch at PA $45$\arcdeg)
and non-barred (hatch at PA $-45$\arcdeg)
Im galaxies showing the influence of bars on star
formation and surface brightness.
The star formation rate (SFR$_D$; from Hunter \& Elmegreen 2004)
is the integrated rate normalized to the area $\pi R_D^2$.
The quantity $\mu_D^V$ is the average V-band surface brightness within
R$_D^V$.
R$_{HII}$/R$_D^V$ is the radius at which the furthest \protect\HII\ region
is found in the disk, relative to R$_D^V$.
The barred Ims have higher
star formation rates, higher central surface brightnesses, and
larger radial extents of the \protect\HII\ regions than the non-barred Ims.
This figure appears in color in the electronic version of the {\it Astrophysical
Journal}.
\label{fig-barsIm}}
\end{figure}

\begin{figure}
\caption{Number distribution of galaxies in the sample with
a given projected minor-to-major axis ratio $b/a$.
The vertical dashed line marks the median value in $b/a$ for
the Im sample.
\label{fig-ba}}
\end{figure}

\clearpage

\begin{figure}
\epsscale{0.6}
\caption{A model for the distribution of the apparent ratio of
axes for randomly projected triaxial galaxies with an intrinsic
width to length ratio uniformly distributed between 0.7 and 1,
and an intrinsic thickness to length ratio uniformly distributed
between 0.29 and 0.67.
\label{fig-wl2}}
\end{figure}

\begin{figure}
\epsscale{0.6}
\caption{(top) Average surface brightness $\mu_{2.5R_D}$ within a radius of 2.5R$_D^V$ in
V-band as a function of the minor-to-major axis ratio of the galaxy.
The normalizing area is the area of the ellipse, not the circular area
in the plane of the galaxy as used for $\mu_{2.5D}$ in Table 5.
The lines indicate median values in $b/a$ bins of 0.1 width.
There is no correlation but the galaxies with the faintest
surface brightnesses tend to lie near the peak in the overall distribution of
$b/a$. (bottom) Absolute V-band magnitude versus axial ratio, showing a general
brightening of all galaxy types for increasing $b/a$.
This figure appears in color in the electronic version of the {\it Astrophysical
Journal}.
\label{fig-sblongrd_bamv}}
\end{figure}

\clearpage

\begin{figure}
\epsscale{0.6}
\caption{(top) Distribution of 2.5R$_D^V$
in kpc is shown versus axial ratio for the three galaxy types.
The lines indicate median values in $b/a$ bins of 0.1 width.
There is no significant trend. (bottom) Integrated color versus axial
ratio, with a trend toward redder colors for more circular projected shapes
in the case of BCD galaxies.
This figure appears in color in the electronic version of the {\it Astrophysical
Journal}.
\label{fig-babmva_longrd}}
\end{figure}

\begin{figure}
\caption{ Integrated UBV colors of our survey
galaxies. Average colors are shown for spiral galaxies (de
Vaucouleurs \& de Vaucouleurs 1972). The UBV colors are corrected
for reddening using the foreground reddening of Burstein \& Heiles
(1984), an assumed internal reddening of E(B$-$V)$_i=$0.05, and
the reddening law of Cardelli \protect\et\ (1989).
\label{fig-ubv}}
\end{figure}

\begin{figure}
\caption{ Integrated JHK colors of our survey
galaxies corrected for reddening. For comparison, several spirals,
an \protect\HII\ region, and a globular cluster that we observed are shown
as well. The average JHK colors of a sample of Im galaxies
observed by Fioc \& Rocca-Volmerange (1999) falls in the middle of
this diagram (J$-$H$=$0.6, H$-$K$=$0.2).
The error bars for the spiral and globular cluster colors are the size of
the plotted points and are not drawn.
\label{fig-jhk}}
\end{figure}

\begin{figure}
\caption{ Integrated VJH colors of our survey
galaxies corrected for reddening.
\label{fig-vjh}}
\end{figure}

\clearpage

\begin{figure}
\caption{Azimuthally-averaged colors as a function of radius
in the Sm galaxy NGC 2552. Solid lines are fits to color gradients.
This is an example of complex color profiles, seen in U$-$B and
B$-$V, and the uncorrelated nature of the different colors.
\label{fig-colorsvar}}
\end{figure}

\begin{figure}
\caption{B$-$V color gradients normalized to the disk scale-length
R$_D^V$ plotted against the integrated absolute V magnitude of the
galaxy (top), the central V-band surface brightness from the
exponential disk fit $\mu_0^V$ (middle), and
the disk scale-length R$_D^V$ (bottom).
Black symbols are the second part to two-part color profiles.
\label{fig-colorprofs}}
\end{figure}

\begin{figure}
\caption{Colors for integrated stellar
populations from the Bruzual \& Charlot (2003) models. Decaying
single population models are on the left and models with
exponentially decaying star formation rates beginning 10 Gyr ago
are on the right. The decay times are on the abscissa.
\label{fig-bruzual}}
\end{figure}

\begin{figure}
\epsscale{0.8}
\caption{B/V images of NGC 2366 (top) and NGC 4449 (bottom).
Black denotes a higher B/V ratio, or bluer colors.
In these two galaxies there is a blue ridge
that crosses the rectangular or boxy part of the galaxy from
one corner to the other.
North is at the top and East to the left.
\label{fig-cross}}
\end{figure}

\begin{figure}
\caption{Number distribution of galaxies in the sample with
a given scale-length R$_D^V$ from fits to the
V-band radial surface photometry profiles.
The vertical dashed line denotes the median value of R$_D^V$
for the Im sample.
\label{fig-histrdv}}
\end{figure}

\clearpage

\begin{figure}
\caption{Number distribution of galaxies in the sample with
a given central surface brightness
$\mu_0^V$ from fits to the V-band radial surface photometry profiles.
The vertical dashed line denotes the median value of $\mu_0^V$
for the Im sample.
\label{fig-histmu0v}}
\end{figure}

\begin{figure}
\caption{V-band scale-length R$_D^V$ versus J-band scale-length R$_D^J$
from fits to the V-band and J-band surface photometry profiles.
The solid line denotes equal values.
\label{fig-rdjrdv}}
\end{figure}

\begin{figure}
\epsscale{0.6}
\caption{Scale-length R$_D^V$ as a function of
integrated M$_{V_0}$ corrected for reddening (left)
and central V-band surface brightness (right).
The ``de Jong spirals''
are data from de Jong (1996a) and de Jong \& van der Kruit (1994).
The de Jong data were originally corrected for Galactic extinction, and
we have corrected for internal extinction with an E(B$-$V)$_i$
of 0.3 mag for each galaxy and no separate correction to face-on orientation.
\label{fig-mvrdv}}
\end{figure}

\begin{figure}
\caption{(Top) Central surface brightness in
V-band and (bottom) V-band disk scale-length
versus galaxy type. The ``de Jong spirals''
are data from de Jong (1996a) and de Jong \& van der Kruit (1994).
The de Jong data were originally corrected for Galactic extinction.
We have corrected for internal extinction with an E(B$-$V)$_i$
of 0.3 mag for each galaxy.
We plot our BCDs as galaxy type 11 in order to separate them from
the Im systems.
The horizontal lines through the Im, BCD, and Sm samples are
the median values of $\mu_0^V$ and R$_D^V$ for our samples.
\label{fig-galtype}}
\end{figure}

\begin{figure}
\caption{Central surface brightness in
V-band versus integrated galaxy magnitude.
The central surface brightness is for the primary exponential
in those cases where the surface brightness is complex.
The ``de Jong spirals''
are data from de Jong (1996a) and de Jong \& van der Kruit (1994).
The de Jong data were originally corrected for Galactic extinction.
We have corrected for internal extinction with an E(B$-$V)$_i$
of 0.3 mag for each galaxy.
The average central
surface brightness
for spiral galaxies, from Freeman (1970), is indicated by the dashed line.
Freeman's $\mu_0^B=21.65$ mag arcsec$^{-2}$ 
has been corrected for reddening in the
same fashion that de Jong's spirals have been corrected
(E(B$-$V)$_i=0.3$)
and a (B$-$V)$_0=0.6$ is used to convert $\mu_0^B$ to $\mu_0^V=19.86$ mag
arcsec$^{-2}$.
This figure appears in color in the electronic version of the {\it Astrophysical
Journal}.
\label{fig-mvmu0_primary}}
\end{figure}

\clearpage

\begin{figure}
\epsscale{0.7}
\caption{V-band surface brightness and B$-$V color profiles for
Im galaxies with double exponential disks.
Each galaxy is offset vertically for clarity and plotted with the
same line color in the upper and bottom panels.
The major ticmarks on the ordinate of the top panel
correspond to 2 mag arcsec$^{-2}$.
The large ticmarks in the bottom panel correspond to 0.2 mag
with red increasing towards the top.
Galaxies that are identified in Table \protect\ref{tab-stuff}
as barred are marked with a ``B'' and those that are identified
as having a peculiar morphology are marked with an ``M.''
Profiles are labeled with the name of the galaxy, where DDO has been
shortened to D, NGC to N, IC to I, and UGC to U.
\label{fig-steepouter_im}}
\end{figure}

\begin{figure}
\caption{As for Figure \protect\ref{fig-steepouter_im} but for the Sm
galaxies with double exponential disks.
\label{fig-steepouter_sm}}
\end{figure}

\begin{figure}
\caption{As for Figure \protect\ref{fig-steepouter_im} but for
the galaxies with nearly uniform inner disks.
\label{fig-flatinner}}
\end{figure}

\clearpage

\begin{figure}
\caption{As for Figure \protect\ref{fig-steepouter_im} but for
the galaxy with a flatter outer exponential disk.
\label{fig-flatouter}}
\end{figure}

\begin{figure}
\caption{As for Figure
\protect\ref{fig-steepouter_im} but for Im galaxies with shallower
outer exponential profiles. IC 10 and NGC 3738 are very similar
to BCD-type systems, shown in Figure
\protect\ref{fig-shallowouter_bcd}, and like the BCDs have very
blue inner disks due to intense, centrally concentrated star formation.
This figure appears in color in the electronic version of the {\it Astrophysical
Journal}.
\label{fig-shallowouter_im}}
\end{figure}

\begin{figure}
\caption{As for Figure \protect\ref{fig-steepouter_im} but for
BCD galaxies with shallower outer exponential profiles.
All but Mrk 600 have
very blue inner disks, indicative of central starbursts.
\label{fig-shallowouter_bcd}}
\end{figure}

\clearpage

\begin{figure}
\caption{Ratio of the break radius R$_{Br}$
to the main exponential scale-length R$_D^V$
versus the V-band
central surface brightness $\mu_0^V$ for galaxies in our sample
that have a double exponential surface brightness profile.
Also shown are the four spirals from de Grijs et al.\ (2001)
which have $\mu_0^V$ (other studies used different pass-bands). Two
of the de Grijs et al.\ spirals have break radii that are significantly
different on the two sides of the galaxy. We have plotted these
galaxies with the larger of the two radii.
\label{fig-breakradii5}}
\end{figure}

\clearpage

\begin{figure}
\epsscale{0.7}
\caption{V-band and H$\alpha$ surface brightness profiles
are plotted with \protect\HI\ surface density profiles for survey galaxies with \protect\HI\
interferometric data in the literature.
The quantities are plotted in equal log intervals.
The line labelled R$_{25}$ marks the radius at which the B-band surface
brightness profile reaches 25 magnitudes in 1 arcsec$^2$, and
R$_H$ marks the Holmberg radius.
The solid line is a fit to the V-band surface photometry. The other
lines just join the H$\alpha$ and \protect\HI\ surface brightness points,
plotted as the logarithm.
DDO 75, DDO 105, and DDO 168 show breaks in their surface
brightness profiles.
References for the \protect\HI\ are as follows:
DDO 43---Simpson, Hunter, \& Nordgren 2005;
DDO 50---Puche et al.\ 1992;
DDO 53---Nordgren et al. 2003;
DDO 75---Skillman et al.\ 1988;
DDO 88---Simpson, Hunter, \& Knezek 2005;
DDO 105---Broeils 1992;
DDO 154---Carignan \& Beaulieu 1989;
DDO 155---Carignan et al.\ 1990;
DDO 168---Broeils 1992;
IC 1613---Lake \& Skillman 1989;
NGC 2366---Hunter, Elmegreen, \& van Woerden 2001a;
and UGC 199---Hunter \& Wilcots 2002.
\label{fig-withhi}}
\end{figure}



\end{document}